\documentstyle[psfig]{mn}

\title[RQQ environments at $0.5 \leq z \leq 0.8$]{Radio-quiet quasar environments at $0.5\le z \le 0.8$}

\author[Wold et al.\ ]
 {Margrethe Wold,$^{1}$\footnotemark[4]
  Mark Lacy,$^2$
  Per B.~Lilje$^3$ and 
  Stephen Serjeant$^4$\\
  $^1$Stockholm Observatory, SE-133 36 Saltsj{\"o}baden, Sweden \\
  $^2$IGPP, Lawrence Livermore National Laboratory, 7000 East Avenue, 
Livermore, CA~94550, USA, \\ and Department of Physics, University of California, 
1 Shields Avenue, Davis, CA~95616, USA \\
  $^3$Institute of Theoretical Astrophysics, University of Oslo, P.O. Box 1029 Blindern, N-0315 Oslo, Norway \\
  $^4$Astrophysics Group, Imperial College London, Blackett Laboratory, Prince Consort Road, London SW7 2BZ, U.K. \\}

\date{Accepted 0000.
      Received 0000;
      in original form 0000}

\pubyear{2000}

\begin{document}

\maketitle


\begin{abstract}
We have quantified the galaxy environments around a sample
of $0.5 \leq z \leq 0.8$ radio-quiet quasars using the amplitude
of the spatial galaxy--quasar correlation function, $B_{\rm gq}$.
The quasars exist in a wide variety of environments, some sources are
located in clusters as rich as Abell class 1--2 clusters, whereas others
exist in environments comparable to the field. 
We find that on average, the quasars prefer poorer clusters of 
$\approx$ Abell class 0, which suggests that
quasars are biased tracers of mass compared to galaxies.
The mean $B_{\rm gq}$ for the sample is found to be 
indistinguishable from the mean amplitude for a sample
of radio-loud quasars matched in redshift and optical luminosity.
These observations are consistent with recent studies
of the hosts of radio-quiet quasars at low to intermediate redshifts, and 
suggest that the mechanism for the production of powerful radio jets 
in radio-loud quasars 
is controlled by processes deep within the active galactic 
nucleus itself, and is unrelated
to the nature of the hosts or their environments.
\end{abstract}


\section{Introduction}

\label{section:s1}

\footnotetext{\footnotemark[4]Email: wold@astro.su.se}

The study and interpretation of 
the differences and similarities between radio-loud and 
radio-quiet quasars (RLQs and RQQs, hereafter)
has kept astronomers busy for a long time. The  
spectral energy distributions of RLQs and RQQs differ only in points of detail 
at all wavelength bands, apart from the radio.
Both quasar types possess broad emission lines and have a stellar
appearance in even the highest resolution optical images.
They also show the same evolution in number density with a peak 
at $z\sim 2$--3, but the RQQs outnumber the RLQs by a 
factor of 10. RQQs have radio luminosities 2--3 orders of 
magnitude lower than their radio-loud counterparts, and on radio maps
they are compact with typically only a weak, flat-spectrum 
component coincident with the optical nucleus. On the other hand, the RLQs have 
extended lobes of radio emission with hotspots at the
outer edges of the radio structure. 
The radio-emitting lobes are fed by powerful jets 
emerging from a bright, central core, but RQQs
can also have jet-like structures 
(Blundell et al.\ 1996; Blundell \& Beasley 1998),
although with bulk kinetic powers $\sim 10^{3}$ times lower 
than for RLQs (Miller, Rawlings \& 
Saunders 1993). This suggests that both quasar populations 
have jet-producing central engines, but that the efficiency
of the jet production mechanism is very different in the two 
cases.

One way to learn more about how the two quasar types relate, is 
to study their host galaxies and their galactic environment. 
A longstanding belief that RQQs are hosted by 
spiral galaxies and RLQs by ellipticals is now being questioned. Recent studies 
(e.g.\ Dunlop et al.\ 1993; McLeod \& Rieke 1994; Taylor et al.\ 1996; 
Bahcall et al.\ 1997; Boyce et al.\ 1998; 
McLure et al.\ 1999; Hughes et al.\ 2000) have found that 
powerful quasars at $z \la 0.5$, both radio-loud and radio-quiet, 
seem to exist in galaxies with luminosities $>L^{*}$, the luminosity at the
break in the luminosity function. Still, a clear picture has not emerged, with some 
studies claiming a high fraction of disk morphologies amongst the 
radio-quiets (e.g.\ Percival et al.\ 2000), whilst others
suggest that nearly all quasars are in giant ellipticals 
(e.g.\ McLure et al.\ 2000). 
Besides the technical difficulties associated with studying host galaxies,
quasar optical luminosity and selection effects
may affect the results. E.g.\ the perhaps somewhat artificial distinction between
Seyfert galaxies and quasars at $M_{V}=-23$ leads to confusion as to whether 
an active galactic nucleus (AGN) is a quasar or a Seyfert.
Seyfert galaxies are spirals, are found at lower redshifts and have
lower AGN luminosities than RLQs, so very few RLQs exist for comparison.
Selection effects may therefore have caused spirals to 
be preferred as hosts of RQQs and 
giant ellipticals to be preferred as hosts of RLQs.

Quasar environments have also been investigated on much larger scales than those
of typical host galaxies. RLQs at $0.5 \la z \la 0.8$
are often found to be associated with groups or clusters of galaxies 
(Yee \& Green 1987; Ellingson, Yee \& Green 1991a; Wold et al.\ 2000 
(hereafter Paper~I)), with richnesses varying from 
poor groups to Abell Class 1 clusters or richer. This is perhaps not surprising
since giant elliptical galaxies, found to be the hosts of powerful RLQs, 
frequently reside in the centres of galaxy clusters.
But what is the galaxy environment like for RQQs in this redshift range,
and how does it compare with the environment around RLQs 
with comparable AGN luminosities?

In this paper, we try to answer this question by investigating a sample
of RQQs matched in redshift and optical luminosity to a sample
of RLQs analysed in Paper~I. We quantify the galaxy environment using the
amplitude of the spatial galaxy--quasar cross-correlation 
function, $B_{\rm gq}$, often referred to as the `clustering amplitude'. 
The first attempt at comparing the environments around RLQs and 
RQQs using the clustering amplitude was made by Yee \& Green \shortcite{yg84}, 
who found only a marginal difference. Later,
Yee \& Green \shortcite{yg87} did deeper imaging of the same RQQ fields, and  
added more RLQ fields to the sample. They repeated the analysis with 
improved parameters, and obtained an even smaller difference, but due to the
small number of RQQs (seven) in their sample, they were unable to draw any
firm conclusions. 

The problem was later addressed by Ellingson et al.\  \shortcite{eyg91a}, who also 
used the clustering amplitude to quantify the
Mpc-scale environment around a sample of 65 $17.5 < m < 19$ quasars at $0.3 < z < 0.6$, of
which half were radio-quiet. 
They found that the RLQs exist more often in rich galaxy environments than the
RQQs do, which was believed to support the RQQ/spiral 
host--RLQ/elliptical host hypothesis.
Since this study, there has been little work aimed at 
comparing the environments of the two quasar populations 
at moderate redshifts, but at both lower and higher redshifts 
several studies indicate that contrary to the conclusion reached by
Ellingson et al., the galaxy environment around RLQs and RQQs is similar.
But also here, a clear picture has not yet emerged, since several other
studies seem to support Ellingson et al's conclusion.

Some investigators use the clustering amplitude, which makes 
comparison between studies similar to ours easier, whereas other 
studies have other aims, and therefore use different
techniques and methods.
These studies can basically be divided into two categories. 
The first category includes measurements of the {\em angular} 
galaxy--quasar cross-correlation function by cross-correlating 
quasar catalogues with galaxy counts. The aim 
is to establish whether optically and X-ray selected quasars
(i.e.\ mostly RQQs) can be used as unbiased tracers of galaxies.
This is important to know, since the large quasars surveys currently
in progress, like the 2dF \cite{shanks00} and the Sloan Digital
Sky Survey \cite{fan99}, 
will be used to study large-scale structure out to high redshifts,
but require knowledge about how quasars are biased with respect
to galaxies. 
The second category includes more or less qualitative searches for
galaxies in quasar and radio galaxy fields at both high and low
redshifts. 
E.g.\ data obtained for host galaxy studies, at typically
$z \la 0.3$, can be utilized for environment studies, but the
high resolution imaging required to study hosts often results in 
a small field of view and therefore a relatively small sampling radius.

Investigations of the galaxy--quasar angular correlation function
are often concerned with scales larger than typical cluster scales.
At low redshifts, 
Smith, Boyle \& Maddox \shortcite{sbm95} selected 169 ($z<0.3$) 
optically identified quasars from the EMSS (Einstein Medium Sensitivity 
Survey), and cross-correlated
their positions with $B_{J}<20.5$ galaxies from the APM galaxy catalogues in a 
region of radius 0\fdg5 centered on the quasar. 
They found that the amplitude of the
angular galaxy--quasar cross-correlation function was identical to that of the 
APM galaxy angular correlation function, implying that the quasars inhabit regions 
of space similar to that of normal galaxies. 
(At $z=0.3$, 0\fdg5 corresponds to almost 10 Mpc).
In a follow-up study, Smith, Boyle \& Maddox \shortcite{sbm00} 
cross-correlated X-ray selected quasars at 
$0.3 \leq z \leq 0.7$ 
with $V\leq23$ galaxies on CCD images.
They found a 2$\sigma$ excess around the $z \leq 0.5$ quasars 
on scales $<1$ arcmin, but no excess for the $z>0.5$ quasars. 
As an overall
result, they claimed that the galaxy--quasar correlation function 
is consistent with that for faint galaxies.
The quasars in their sample span a wide range in luminosity, 
from $M_{V} \approx -27$ to $-$20 ($H_{0}=50$ km\,s$^{-1}$\,Mpc$^{-1}$), but 
no difference between high and low luminosity sub-samples was observed.

Croom \& Shanks \shortcite{cs99} cross-correlated 150 quasars selected 
both in the optical and in X-rays with $b_{J}<23$ galaxies on AAT
(Anglo-Australian Telescope) photographic plates. 
They found the angular cross-correlation to be marginally
negative, indicative of a small anti-correlation, possibly due to gravitational lensing.
Most of the quasars in their sample lie at redshifts 1.0--2.2,
and they claimed to reach $M^{*}$ at $z\approx1$ and 
$M^{*}-1$ at $z\approx1.5$. Possibly, this is not deep enough, as even
$M^{*}$ might be too shallow into the luminosity function to detect excess
galaxies at $z=1.0$--2.2 \cite{ylc99}.

Searches for galaxies in quasar and radio galaxy fields in connection with
host studies have been conducted by e.g.\ Smith \& Heckman
\shortcite{sh90} who generated catalogues of companion galaxies
within 100$h^{-1}$ Mpc of 31 $z < 0.3$ quasars and 35 powerful radio galaxies.
They found that RQQs only have half the number of nearby, bright
companions as RLQs and radio galaxies at the 95 per cent confidence level. 
Somewhat inconsistent with Smith \& Heckman's result is the result obtained by 
Fisher et al.\ \shortcite{fisher96} who found no significant difference in the 
environments around optically luminous RLQs (six) and RQQs (14)
at $z \la 0.3$.  
Supporting Fisher et al's result is the recent study by McLure \& Dunlop
\shortcite{md00}, who quantified the environments around 44 powerful,
$z\approx0.2$ AGN, of which 13 were RLQs and 21 were RQQs (their sample
includes Fisher et al's quasars). They found the mean clustering 
amplitude for the two quasar samples to be consistent with each other,
and corresponding to Abell richness class 0 clusters.
Dunlop et al.\ \shortcite{dunlop93}, studying quasar hosts in the 
near-infrared, also discuss companion galaxies. Their sample consisted of
14 RLQs and 14 RQQs at $z<0.35$ matched in redshift and
$V$ magnitude. They found that the surface density of bright ($K > 18$)
companion galaxies was a factor of $\approx 2.5$ higher than in the field, 
and the excess was similar for both RLQs and RQQs. 

The high redshift studies of quasar environments typically concentrate
at $z \ga 1$. Quasars (and radio galaxies) are used as pointers to 
high redshift galaxies in order to learn about galaxy formation and
evolution and to search for high redshift (proto)clusters. Galaxies,
and even clusters, are certainly found at $z \ga 1$ using this method
(e.g.\ Dickinson 1994;
Deltorn et al.\ 1997; Hall \& Green 1998; Tanaka et al.\ 2000, Chapman, 
McCarthy \& Persson 2000; Best 2000).

Hintzen, Romanishin \& Valdes \shortcite{hrv91} and 
Boyle \& Couch \shortcite{bc93} investigated 
$0.9 < z < 1.5$ RLQs and RQQs, respectively.
Hintzen et al.\ found a significant galaxy excess 
within a 15 arcsec radius ($\approx160$ kpc) centered on the 
quasars in their sample,
whereas Boyle \& Couch \shortcite{bc93} found
no evidence for excess.
However, Boyle \& Couch's RQQs are 20 times
fainter in the optical than the RLQs studied by Hintzen et al., 
and Boyle \& Couch discuss this as a possible cause 
for the discrepancy. 

In contrast to this, Hutchings, Crampton \& Johnson \shortcite{hcj95} 
found on average a $>5$$\sigma$ excess of galaxies around both
RLQs and RQQs (14 in total) at $z \approx 1.1$. They saw no 
significant difference between RLQ and RQQ environments, which was
found to consist of compact groups of starbursting galaxies.
Also, Hutchings \shortcite{hutchings95a}, who looked at galaxy companions
to seven $z=2.3$ quasars (only one RLQ) in the $R$-band
found an overall $>5$$\sigma$ excess in a 
$\sim 1$ arcmin region around the quasars.
However, Hutchings et al.\ \shortcite{hutchings99}, using near-infrared
imaging of both RLQs and RQQs at $0.9 < z < 4.2$, note that the RQQs
occupy poorer environments than the RLQs, but do not draw any firm
conclusions due to the small and still incomplete sample.
Teplitz, McLean \& Malkan \shortcite{tmm99} also report that the 
number counts of galaxies in RQQ fields at 
$0.95<z<1.5$ in $J$- and $K'$-band is comparable to the field 
counts, but their data is not very deep (only reaching 
$\approx M^{*}$ at $z \approx 1$). 

In summary, although the results seem to differ, 
the consensus about hosts and environments of 
RQQs -- that they are typically late-type galaxies in poor 
environments -- is being increasingly questioned. 
Several recent observations at high and low redshifts seem to indicate that
RQQs live in richer than average galaxy environment, and powerful RQQs
at $z \la 0.5$ are found in luminous, possibly early-type, galaxies.

Apart from Ellingson et al's \shortcite{eyg91a} study at $0.3 < z < 0.6$,
there have been no attempts to investigate what environments RQQs prefer
at intermediate redshifts. We have therefore obtained data on a sample
of 20 RQQs at $0.5 \leq z \leq 0.8$, and quantified the galaxy richness
in the quasar fields using a similar analysis as Ellingson et al.
The sample is described in Section~\ref{section:s2}, where we also
discuss the radio loudness of some of the quasars.
In Section~\ref{section:s3}, we describe the observations and our 
observing strategy. The control fields used in the analysis are also
commented upon here, especially how the galaxy counts vary with
galactic latitude. The analysis using the clustering amplitude is 
briefly overviewed in Section~\ref{section:s4}, before the 
results are presented in Section~\ref{section:s5}.
In Section~\ref{section:s51} we examine the colours
and the radial distribution of the galaxies in the quasar fields.
Using the clustering amplitude allows us to compare our results with 
some of the above mentioned studies, and this is done in 
Section~\ref{section:s6}, and in Section~\ref{section:s7}
we are able to directly and consistently compare with the sample
of RLQs that was presented 
in Paper~I. Finally, we discuss the results in 
Section~\ref{section:s8}, and draw the conclusions in Section~\ref{section:s9}.

Our assumed cosmology has $H_{0}=50$ km\,s$^{-1}$\,Mpc$^{-1}$, 
$\Omega_{0}=1$ and $\Lambda=0$.

\begin{table*}
\begin{minipage}{13.5cm}

\caption{The quasar sample and log of observations. All images 
were obtained using the HiRAC equipped with a 1k SiTe CCD 
with pixel scale 0.176 arcsec, unless otherwise is specified 
in footnotes. In the 6th column, we list the FWHM of the seeing in 
the combined image (in the longest wavelength filter when 
there are observations in two filters).}

\begin{tabular}{llllllll}
Quasar   & $z$  & Filter & Exp. time  & Date  &  Seeing & Remarks\\  
       &           &   & (s)  &           & (arcsec) & \\
       &           &     &           &       &    & \\
LBQS 0007$-$0003  & 0.698 & $I$      
& 4$\times$600 & 96/07/23 & 1.1 & photometric\\

LBQS 0020$-$0300  & 0.580 & $V$, $R$ 
& 4$\times$600 & 96/07/25 & 0.8 & dust \\

BFSP 12344$-$0021\footnotemark[1]\footnotetext{\footnotemark[1]Imaged using ALFOSC with a 2k Loral CCD with pixel scale 0.189 arcsec.} & 0.753 & $I$ 
& 5$\times$300 & 98/04/28 & 0.7 & variable transparency \\
       &      &      $I$    
&   6$\times$300\footnotemark[3]  &  00/01/03 &   &  photometric\\

BFSP 12355$+$0107\footnotemark[2]\footnotetext{\footnotemark[2]Imaged with the HiRAC camera, but with a 2k Loral CCD with pixel scale 0.11 arcsec.} & 0.720 & $I$      
& 9$\times$300 & 97/05/14 & 0.5 & photometric \\

BFSP 12364$-$0053\footnotemark[1] & 0.727 & $I$      
& 8$\times$300 & 99/05/19,23 & 0.9 & variable transparency\\
       &         &      $I$   
& 1$\times$300\footnotemark[3]\footnotetext{\footnotemark[3]Exposures taken during photometric conditions in order to calibrate the other images obtained under less photometric conditions.}  &  00/01/03 &   & photometric \\

BQS 1333$+$176\footnotemark[1]  & 0.554 & $R$      
& 4$\times$600 & 00/01/03 & 0.7 & photometric \\

BFSP 15152$+$0244\footnotemark[2] & 0.608 & $R$      
& 4$\times$600 & 97/05/14 & 0.6 & photometric\\

BFSP 15196$+$0220   & 0.742 & $I$      
& 4$\times$600 & 96/07/24 & 0.8 & dust \\

BFSP 15199$+$0247  & 0.768 & $R$, $I$ 
& 5$\times$600 & 96/07/23 & 0.6 & photometric\\

BFSP 15202$+$0236  & 0.765 & $R$, $I$ 
& 4$\times$600 & 96/07/22 & 0.6 & photometric\\

BQS  1538$+$477    & 0.770 & $R$, $I$ 
& 4$\times$600 & 96/07/21 & 0.6 & photometric\\

BFSP 22011$-$1857 & 0.615 & $V$, $R$ 
& 4$\times$600 & 96/07/21 & 1.0 & photometric\\

LBQS 2235$+$0054  & 0.529 & $V$, $R$ 
& 4$\times$600 & 96/07/21 & 1.0 & photometric\\

LBQS 2238$+$0133  & 0.714 & $R$, $I$ 
& 4$\times$600 & 96/07/22 & 0.7 & photometric\\

LBQS 2239$-$0055 & 0.680 & $R$, $I$ 
& 4$\times$600 & 96/07/23 & 0.9 & photometric \\  

LBQS 2245$-$0055  & 0.801 & $R$, $I$ 
& 5$\times$600 & 96/07/24 & 1.3 & dust \\  

LBQS 2348$+$0210  & 0.504 & $V$, $R$ 
& 4$\times$600 & 96/07/24 & 1.3 & dust\\

LBQS 2348$+$0148 & 0.749 & $I$      
& 4$\times$600 & 96/07/25 & 0.8 & dust\\

LBQS 2350$-$0012  & 0.561 & $V$, $R$ 
& 4$\times$600 & 96/07/23 & 0.9 & photometric\\

LBQS 2353$-$0153  & 0.672 & $R$, $I$ 
& 4$\times$600 & 96/07/22 & 0.6 & photometric\\
\end{tabular}
\end{minipage}
\label{table:table1}
\end{table*}

\begin{table*}
\begin{minipage}{13.0cm}

\caption{Radio fluxes at 1.4 GHz from NVSS and FIRST image cutouts. Quasars with radio detections
in NVSS or FIRST (or both) are marked with an $R$ in the first column. For the other quasars, we
give upper limits that correspond to 3$\sigma$ detections. Rest frame fluxes at 
5 GHz and 2500 {\AA} are denoted $S_{\rm 5 GHz}$ and 
$S_{2500}$, and their ratio is $R_{\rm 5 GHz}$. The radio power
is denoted $P_{\rm 5 GHz}$, and $\alpha_r$ is the radio spectral 
index between 1.4 and 5 GHz.}

\begin{tabular}{llllllll}
Quasar & NVSS  & FIRST & $S_{\rm 5 GHz}$ &$\alpha_r$& $S_{\rm 2500}$ & $R_{\rm 5 GHz}$ & $P_{\rm 5 GHz}$ \\
       & (mJy) & (mJy) & (mJy)           &          & (mJy)          &                 & W\,Hz$^{-1}$    \\
       &       &       &                 &          &                &                 &           \\
LBQS 0007$-$0003  & $<$ 1.5 & $<$ 0.5 & $<$0.61     & 0.50 & 0.10 & $<$ 6.1 & $<$ 1.6$\times$10$^{24}$ \\

LBQS 0020$-$0300  & $<$ 1.8 & no map  & $<$0.76     & 0.50 & 0.21 & $<$ 3.6   & $<$ 1.4$\times$10$^{24}$ \\   

BFSP 12344$-$0021$^{\rm R}$ & 4.8     & 5.2     & 1.92        & 0.50 & 0.03 & 64.0 & 6.1$\times$10$^{24}$ \\

BFSP 12355$+$0107 & confused& $<$ 0.5 & $<$0.20     & 0.50 & 0.02 & $<$ 10.0  & $<$ 5.8$\times$10$^{23}$ \\

BFSP 12364$-$0053 & $<$ 1.5 & $<$ 0.5 & $<$0.60     & 0.50 & 0.01 & $<$ 60.0  & $<$ 1.8$\times$10$^{24}$\\

BQS 1333$+$176$^{\rm R}$    & 37.6    & no map  & 18.54       & 0.32\footnotemark[1]\footnotetext{\footnotemark[1]Spectral index found between flux at 1.4 GHz and 
flux measured at 5 GHz by Kellermann et al.\ (1989)}  & 0.60 & 30.9 & 3.0$\times$10$^{25}$ \\

BFSP 15152$+$0244 & $<$ 1.5 & $<$ 0.5 & $<$0.63     & 0.50 & 0.02 & $<$ 31.5  & $<$ 1.3$\times$10$^{24}$ \\

BFSP 15196$+$0220 & $<$ 1.5 & $<$ 0.5 & $<$0.60     & 0.50 & 0.03 & $<$ 20.0  & $<$ 1.9$\times$10$^{24}$ \\

BFSP 15199$+$0247 & $<$ 1.5 & $<$ 0.5 & $<$0.60     & 0.50 & 0.03 & $<$ 20.0  & $<$ 2.0$\times$10$^{24}$ \\

BFSP 15202$+$0236$^{\rm R}$ & $<$ 1.5 & 1.0     & 0.40        & 0.50 & 0.01 & 40.0 & 1.3$\times$10$^{24}$ \\

BQS 1538$+$447$^{\rm R}$ & 95.2    & 85.4    & 26.28       & 1.02\footnotemark[1] & 1.00 & 26.3 & 8.8$\times$10$^{25}$ \\

BFSP 22011$-$1857 & $<$ 1.5 & no map  & $<$0.62     & 0.50 & 0.05 & $<$ 12.4  & $<$ 1.3$\times$10$^{24}$ \\

LBQS 2235$+$0054$^{\rm R}$ & $<$ 1.5 & 0.83    & 0.94        & $-$0.65\footnotemark[2]\footnotetext{\footnotemark[2]Spectral index between 1.4 GHz flux and flux measured at 8.4 GHz by Hooper et al.\ (1996)} & 0.02 & 47.0 & 1.4$\times$10$^{24}$ \\ 

LBQS 2238$+$0133  & $<$ 1.5 & no map  & $<$0.61     & 0.50 & 0.22 & $<$ 2.8  & $<$ 1.7$\times$10$^{24}$ \\

LBQS 2239$-$0055  & $<$ 1.5 & $<$ 0.5 & $<$0.61     & 0.50 & 0.12 & $<$ 5.1 & $<$ 1.6$\times$10$^{24}$ \\

LBQS 2245$-$0055$^{\rm R}$  & $<$ 1.5 & 1.3     & 0.51        & 0.50 & 0.30 & 1.7  & 1.9$\times$10$^{24}$  \\

LBQS 2348$+$0210$^{\rm R}$  & 41.8    & no map  & 25.05       & 0.12\footnotemark[2] & 0.17 & 143.4 & 3.3$\times$10$^{25}$ \\

LBQS 2348$+$0148  & $<$ 1.5 & no map  & $<$0.60     & 0.50 & 0.08 & $<$ 7.5  & $<$ 1.9$\times$10$^{24}$ \\

LBQS 2350$-$0012  & $<$ 1.5 & $<$ 0.5 & $<$0.64     & 0.50 & 0.32 & $<$ 2.0  & $<$ 1.1$\times$10$^{24}$ \\

LBQS 2353$-$0153  & $<$ 1.5 & $<$ 0.5 & $<$0.61     & 0.50 & 0.57 & $<$ 1.1 & $<$ 1.5$\times$10$^{24}$ \\
\end{tabular}
\label{table:table2}
\end{minipage}
\end{table*}


\section{The sample}

\label{section:s2}

The RQQs were selected from three optical surveys with different 
flux limits in order to cover a wide range in AGN luminosity 
($B$-band) within the given redshift range. 
Eight of the quasars are from the faint ($B \la 21$) Durham/AAT UVX survey 
of Boyle et al.\ \shortcite{bfsp90} (designated `BFSP')
and ten are from the intermediate 
luminosity ($16 \la b_{J} \la 18.9$)
Large Bright Quasar Survey (LBQS) by Hewett, Foltz \& Chaffee \shortcite{hfc95}.
There are also two high-luminosity quasars in the sample,
selected from the $B \la 16.6$ Bright Quasar Survey (BQS)
by Schmidt \& Green \shortcite{sg83}. The sample is listed in Table~\ref{table:table1} 
along with information about the observations.

The BQS and the Durham/AAT UVX survey are based on the UV-excess 
selection technique which finds $z<2.2$ quasars with stellar-like appearance
and excess emission in the UV. The LBQS is an objective prism 
survey, from which quasar candidates are essentially selected by their blue
colour. The LBQS and the Durham/AAT UVX survey are complete,
but the BQS survey has been found to be incomplete by
a factor of 3.4 \cite{goldschmidt92}, possibly due to photometry errors.
Also, the BQS is known to have an anomalously
high radio-loud fraction at bright optical magnitudes,
$M_{B} < -24$, compared to other optically selected
samples \cite{lafranca94}, and Hooper et al.\ \shortcite{hooper96} 
find that the radio-loud fraction 
at faint optical luminosities is lower in the BQS
than in the LBQS.
We have checked that inclusion of the two BQS quasars does not affect our results.

The RLQ sample that we compare with in Section~\ref{section:s7} 
consists of 21 steep-spectrum RLQs at $0.5 \leq z \leq 0.8$ covering
the wide radio luminosity range 
$23.8 \leq \log\left(L_{\rm 408 MHz}/{\rm W}\,{\rm Hz}^{-1}\,{\rm sr}^{-1}\right) \leq 26.7$.
This sample was drawn from two different radio/optical flux-limited
samples, the Molonglo/APM Quasar Survey (Serjeant 1996; Maddox et al., in preparation;
Serjeant et al., in preparation) and the 7C Quasar survey \cite{riley99}.
For details about the RLQ sample, we refer to Paper~I.

In order to check if some of the RQQs were detected at radio wavelengths, 
we extracted image cutouts at the quasar positions from the 1.4 GHz NVSS\footnote{NRAO/VLA
Sky Survey} \cite{nvss98} and the VLA FIRST\footnote{Faint Images of the Radio Sky
at Twenty-centimeters} survey
\cite{bwh95} and examined them in 
{\sc aips}. We found seven quasars that were detected in either 
both or just one of the surveys and list the radio fluxes of these in 
Table~\ref{table:table2}.
For the other quasars, we set a 3$\sigma$ detection
as upper limit since the 1$\sigma$ noise level for the
NVSS and FIRST are 0.5 and 0.17 mJy\,beam$^{-1}$, 
respectively.

The quasars with radio detections are the two BQS quasars, 
three of the LBQS quasars and two of the faint BFSP quasars, and we have
marked these with `$R$' in Table~\ref{table:table2}. 
One of these quasars, BQS 1538$+$447, looks resolved on the 
FIRST map and appears to have FRI morphology. 

Can these seven (out of 20) radio-detected quasars be classified as RQQs? The distinction
between RLQs and RQQs is not clear-cut, 
and different definitions exist. Indeed it seems likely that there is a continuum
in radio-loudness, with RLQs representing the extreme of the distribution
in radio luminosity at a given absolute magnitude \cite{white00}.
Nevertheless, a distinction is often made based on radio luminosity
and/or the ratio of radio to optical flux density, $R$.
According to the radio luminosity
criterion, a quasar is radio-loud if 
$P_{\rm 5 GHz} > 10^{24}$ W\,Hz$^{-1}$\,sr$^{-1}$,
or $P_{\rm 5 GHz} > 1.3\times 10^{25}$ W\,Hz$^{-1}$, whereas 
according to the criterion used by e.g.\ Kellermann et al.\ 
\shortcite{kellermann89}, a quasar is radio-loud if 
$R \ga 10$. 
A much used definition of $R$ is the ratio of the rest 
frame flux at 5 GHz to the $B$-band flux, and here we choose
Sramek \& Weedman's \shortcite{sw80} definition, 
$R=S_{\rm 5 GHz}/S_{2500}$, which uses
the flux density at rest frame 5 GHz and 2500 {\AA} 
(corresponding to $B$-band when $z\sim 0.7$).
The usefulness of the $R$-parameter is not clear, e.g.\ 
Goldschmidt et al.\ \shortcite{goldschmidt99} argue that
$R$ is physically meaningful only if the radio and optical
luminosity is linearly correlated, and that there
are various reasons that this might not be the case
(Goldschmidt et al.\ and references therein).

To calculate the $R$-parameter, we assumed that both radio and optical 
flux density scales as 
$S_{\nu} \propto \nu^{-\alpha}$, and computed rest frame
flux densities at 5 GHz and 2500 {\AA}.
The rest frame flux density at 5 GHz was found according to 
\[
S_{\rm 5 GHz}=S_{\rm 1.4 GHz}\left(\frac{5}{1.4}\right)^{-\alpha}\left(1+z\right)^{\alpha -1},
\]
\noindent
where we used the NVSS flux as $S_{\rm 1.4 GHz}$ 
when available, since this is most sensitive to extended structures.
The radio power was also calculated,
\[
P_{\rm 5 GHz}=4 \pi d_{\rm L}^{2}\,S_{\rm 5 GHz}, 
\]
\noindent
where $d_{L}$ is the luminosity distance to the quasar.
The optical flux density at rest frame 2500 {\AA} was found
from the $m_{B}$ magnitudes of the quasars (determined
from the images, see details in Section~\ref{section:s3}) using the relation
\[
S_{2500}=S_{B}\left(\frac{4400}{2500}\right)^{-\alpha}\left(1+z\right)^{\alpha -1},
\]
\noindent
where $\log S_{B}=-22.35 - 0.4 m_{B}$ \cite{longair81}.
In Table~\ref{table:table2}, we list the rest frame radio and
optical flux densities, the $R_{\rm 5 GHz}$ parameter and the radio
power.

In the above calculations, we assumed a spectral index of 
$\alpha = 0.5$ 
for both the radio and the optical spectra, but for four of the 
quasars we estimated radio spectral indices
by extrapolating between the NVSS/FIRST 
flux densities and flux densities measured 
at 5 GHz by Kellermann et al.\ \shortcite{kellermann89} or at 8.4 GHz
by Hooper et al.\ \shortcite{hooper96}, denoted $\alpha_{\rm r}$
in Table~\ref{table:table2}.
We found that BQS 1538$+$477 has a steep radio spectrum 
with $\alpha^{5}_{1.4}=1.02$, but the spectrum flattens at high frequency,
with $\alpha^{10}_{8.7}=-0.19$ (Falcke, Sherwood \& Patnaik 1996), so it is
therefore 
likely that the bright core is flat spectrum and possibly variable, but the 
extended emission is steep-spectrum.
The spectrum of LBQS 2235$+$0054 is somewhat unusual with 
$\alpha_{1.4}^{8.4}=-0.65$, i.e.\ the flux density rises with 
increasing frequency. 
This could be due to synchrotron self-absorption at low frequencies or perhaps
thermal absorption.

As seen in Table~\ref{table:table2}, six of the seven radio-detected quasars 
have $R > 10$, and classify as radio-loud according
to Kellermann et al's \shortcite{kellermann89} criterion, but
only three of these, the two BQS quasars and
LBQS 2348$+$0210, have radio luminosities greater 
than 1.3$\times$10$^{25}$ W\,Hz$^{-1}$ and qualify both criteria.

Miller et al.\ \shortcite{mrs93}, 
in a study of BQS quasars, found a group of quasars 
with radio luminosities at 4.8 GHz in the range
$\sim 1.3$--$4.0 \times 10^{24}$ W\,Hz$^{-1}$ 
($H_{0}=50$ km\,s$^{-1}$\,Mpc$^{-1}$, $q_{0}=0$) 
that did not tightly follow the emission 
line--radio luminosity correlation, and termed these
`radio-intermediate quasars'.
It was suggested that they could be Doppler-boosted 
RQQs with Lorentz factors of $\sim 5$. 
In terms of the $R$-parameter, it is common to classify 
quasars with $1 < R_{\rm 5 GHz}< 100$ as radio-intermediate 
and quasars with $R_{\rm 5 GHz}\leq 1$ 
as radio-quiet. According to this, all but one
of the seven radio-detected quasars are radio-intermediate, the exception being
LBQS 2348$+$0210 with $R_{\rm 5 GHz}=143.4$.

Falcke et al.\ \shortcite{fsp96} find that the
division line at $R=10$ between RLQs and RQQs
leads to inconsistencies with the Unified Scheme,
since they find a high fraction of flat-spectrum,
core-dominated quasars in the BQS. 
According to orientation-based Unified Schemes these are 
relativistically beamed radio 
sources with extended lobe emission, and should therefore
be rare and have radio flux densities higher than the steep-spectrum
sources. Falcke et al.\ term these radio-intermediate
quasars, and also suggest that they are relativistically boosted
radio-weak quasars. 
They propose to use both the $R$-parameter {\em and} the spectral 
index to divide between RLQs and RQQs, and that the division line
between radio-loud and radio-quiet is set at $R=250$ for 
flat-spectrum sources and at $R=25$ for 
steep-spectrum sources, so that the radio-intermediate quasars 
have $R$-parameters in the range 25--250.

Blundell \& Beasley \shortcite{bb98} also point out that if the 
criterion for radio-loudness
is based on the extended radio emission in quasars, the 
radio-intermediate quasars would classify
as radio-quiet. This would however not be the
case for
BQS 1538$+$447, since its extended radio flux
is 53.9 mJy (total NVSS flux minus
peak FIRST flux), corresponding to 5.0$\times$10$^{25}$ 
W\,Hz$^{-1}$ at 5 GHz.

The seven quasars discussed above clearly have radio powers 
much lower than typical RLQs ($\sim 10^{26}$--10$^{27}$ W\,Hz$^{-1}$),  
and if not radio-quiet, may classify as either radio-intermediate 
or radio-weak. However, we have checked that our results are not affected 
by the presence/absence of these quasars.
 
\begin{figure}
\psfig{file=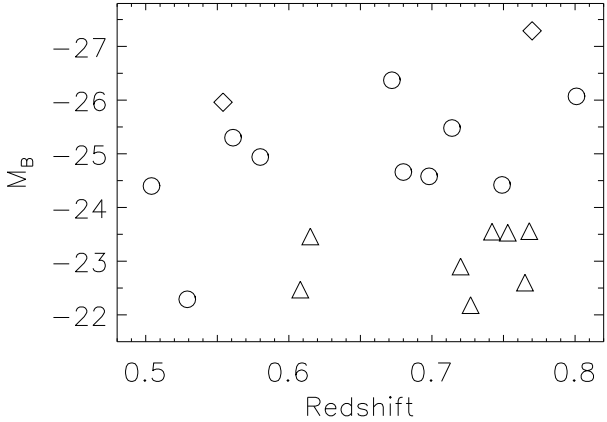}
\caption{Absolute $B$ magnitudes of the quasars as a function of redshift. 
Diamonds: bright BQS quasars, circles: intermediate-luminosity 
LBQS quasars, triangles: faint BFSP quasars.}
\label{fig:fig1}
\end{figure}


\section{Observations and data reduction}

\label{section:s3}

Most of the data were obtained using the High Resolution
Adaptive Camera (HiRAC) at the 2.56-m Nordic Optical Telescope (NOT) during 
1996 July 21--25.
The HiRAC was equipped with a 1k SiTe CCD with a pixel 
scale of 0.176 arcsec, giving a field of view of 3$\times$3 arcmin. 
Two quasar fields were imaged in May 1997, also using the HiRAC, but with a
Loral 2k chip with pixel scale 0.11 arcsec, giving a 3.7$\times$3.7 arcmin field 
of view. 
Images of three more quasar fields were obtained during observing runs at the NOT 
in 1998, 1999 and 2000 using the ALFOSC (Andalucia Faint Object 
Spectrograph and Camera) in imaging mode, equipped with a 2k Loral CCD with pixel scale 0.189 arcsec, and a field of view of 6.5$\times$6.5 arcmin.
The bulk of the data were obtained under photometric conditions, and the seeing FWHM is
less than one arcsec in 15 out of 20 fields.

Since we are attempting to detect galaxies that are physically associated with 
the quasars, the filters were chosen to give preference to early-type
galaxies with strong 4000 {\AA} breaks at the quasar redshifts.  
The spectral energy distribution of early-type galaxies is characterized 
by a strong break at 4000 {\AA} with most of the energy emitted at 
wavelengths longer than 4000 {\AA}. For
an early-type galaxy at $z \geq 0.67$ the 4000 {\AA} break moves from
$R$ into $I$-band, so for the $z \geq 0.67$ quasars we used
$I$-band imaging and for the $z < 0.67$ quasars we chose $R$-band. 
Twelve of the fields were also imaged in two filters, either $V$ and $R$, or $R$ and $I$,
depending on the redshift of the quasar so as to straddle the 4000 {\AA} break.

Typically, the integrations were divided into four exposures of 600 s each, offsetting
the telescope 10 arcsec between each exposure to avoid bad pixels and cosmetic 
defects from falling consistently on the same spots.
The $I$-band images from 1997--2000 (three fields) have a prominent fringing pattern
due to interference of night sky lines. We therefore took 9--11$\times$300 s exposures 
of these fields in order to obtain a sufficient number of images to construct a 
fringe frame. The fringe frame was made by taking the median of nine $I$-band 
images taken close in time, and used to flatten the background with
by dividing out the fringes.
Apart from this flat fielding technique to remove fringes, 
the data were reduced as described in Paper~I
using standard {\sc iraf} reduction tasks.

We also observed standard star fields during each night in order to 
calibrate magnitudes to the standard Johnson photometric system. During the 1996 and 1997 runs,
we observed fields by Christian et al.\ \shortcite{christian85}, and for the 
other observing runs we selected standard stars by Landolt \shortcite{landolt92}.

The two fields from 1998 and 1999 (BFSP 12344$-$0021 and BFSP 12364$-$0053)
were imaged as back-up targets during
another programme that required photometric and sub-arcsec seeing conditions, so the transparency 
is somewhat reduced in these frames. We therefore took exposures of 
the same fields under photometric conditions in January 2000 to 
enable a proper photometric calibration.

For photometry and object detection we processed the images
in {\sc focas} (Faint Object Classification and Analysis System, e.g.\ 
Valdes \shortcite{valdes89}), as described in Paper~I.
Each detected object was classified by using a template PSF 
(a 15$\times$15 array) formed from 4-6 selected stars in each 
image, and a record of {\sc focas} total magnitudes and positions
of the detections were made by excluding objects classified as 
`multiple', `long' or  `noise'.
The detections were examined by eye and objects coinciding with
spikes of bright stars were removed from the records.

As described in Paper~I, we performed completeness simulations in the images,
and these simulations show that the data are complete down to 24.0 in $V$, 23.5 in $R$ 
and 23.0 in $I$, with errors of $\pm$0.3 mag at the limits. 

The quasar fields lie at high galactic latitudes 
$46^{\circ}<\left|b\right|<76^{\circ}$ and the galactic
reddening is $E\left(B-V\right)<0.063$ in all fields. We corrected for galactic extinction using an
electronic version of the maps by Burstein \& Heiles \shortcite{bh82} and the Galactic
extinction law by Cardelli, Clayton \& Mathis \shortcite{ccm89}. 

Unless the quasar was saturated in the image, we determined its 
apparent magnitude in either $R$ or $I$, and thereafter
converted to $B$ assuming an optical power-law spectrum, $S_{\nu} \propto \nu^{-\alpha}$, 
with $\alpha=0.5$ and zero points in $B$, $R$ and $I$ as given by Longair \shortcite{longair81}. 
The $B$ magnitudes were thereafter converted to absolute magnitudes by 
assuming a $K$-correction equal to $2.5\left(\alpha-1\right)\log\left(1+z\right)$.

Three of the quasars were saturated in the images (BQS 1333$+$176, BQS 1538$+$477 and 
LBQS 2348$+$0210) so we used apparent $B$ and $B_{J}$ magnitudes as given in 
the catalogues, except for BQS 1333$+$176 for which Yee, Green \& Stockman \shortcite{ygs86} 
have determined an $r$ magnitude.
The $B_{J}$ magnitude for LBQS 2348$+$0210 was converted to $B$ by assuming $B=B_{J}+0.09$
\cite{metcalfe}, and the Gunn $r$ magnitude of BQS 1333$+$176 was first converted to $R$
by assuming $r=R+0.43+0.15\left(B-V\right)$ \cite{kent85} and a $B-V$ colour of 0.3 
(i.e. $B-V$ of a quasar with optical spectral index $\alpha=0.5$), 
and thereafter to $B$ as described above.

Fig.~\ref{fig:fig1} shows the distribution of the quasars in the 
$z$--$M_{B}$ plane, and the $M_B$ magnitudes 
are also listed in Table~\ref{table:table4}.


\subsection{Control fields}

\label{section:s31}

Since we are concerned with investigating if there is an excess of 
galaxies in the quasar fields, we aim to have a good determination of the
background counts. For this purpose, we obtained images
of random fields in the sky at approximately the same 
galactic latitudes as targets in our AGN sample. The control fields were described
in Paper~I, so here it suffices to mention that these fields were imaged 
as part of our AGN environment survey, and therefore have the 
same depth and were obtained in exactly the same manner as
the quasar fields. This is important in order to get a robust estimate
of the galaxy richness in the quasar fields \cite{ylc99}.

In Paper~I, we averaged the counts in the control fields in order to obtain a good
determination of the background counts, and we also found that the 
slopes of the background galaxy number counts agreed well with that found by other
investigators, e.g.\ Smail et al.\ \shortcite{smail95}.
Due to the clustered nature of field galaxies, the errors in the 
galaxy counts are expected to be $\approx 1.3\sqrt{N}$ \cite{ylc99}. However,
the variation in the counts from control field to 
control field is larger than this, presumably
due to the small areas surveyed in each field (typically 3$\times$3 arcmin),
but also 
due to a variation in the counts with galactic latitude, especially in the $I$-band 
control fields. 

\begin{figure*}
\begin{minipage}{17.6cm}
\psfig{file=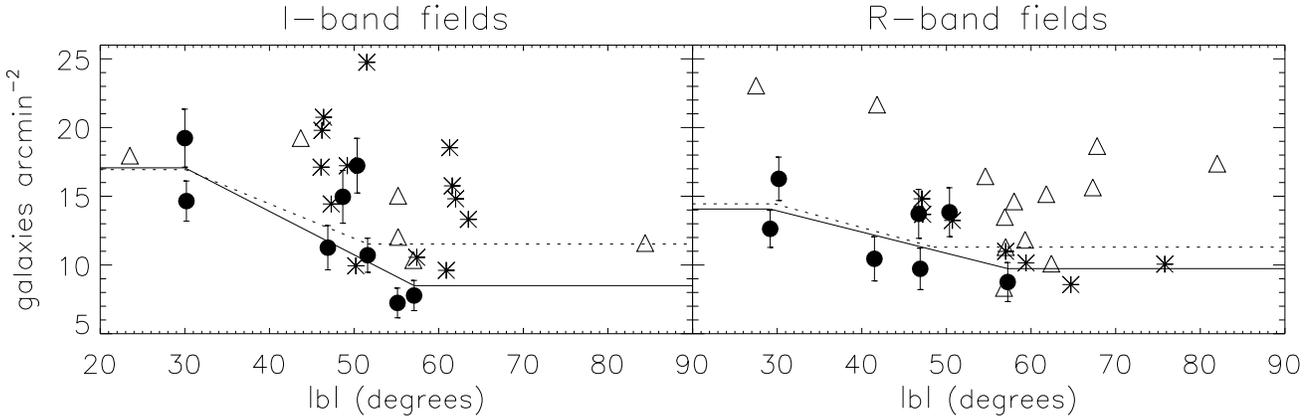}
\caption{The filled circles show the counts of galaxies 
in the $I$- and $R$-band control fields as a function of galactic latitude
(absolute value).
The number of galaxies in the quasar fields are also plotted
using asterisks for the RQQs and triangles for the RLQs.
The solid and dotted lines show the relations for the background counts as a function of galactic latitude, corresponding to Model~1 and Model~2, respectively (see text and Table~\protect\ref{table:table3} for details).}
\label{fig:fig2}
\end{minipage}
\end{figure*}

This can be seen in 
Fig.~\ref{fig:fig2} where we have plotted the number of galaxies brighter than
the completeness limits 
from raw counts in the control fields as a function of galactic latitude,
with 1.3$\sqrt{N}$ error bars. The counts in the quasar
fields within a 0.5 Mpc radius centered on the quasar are also plotted,
and we have included the RLQ fields from Paper~I. The error bars on the 
counts in the quasar fields are intentionally left out for clarity.
It is seen that six of the eight $I$-band control fields lie at galactic
latitudes between 45--60$^{\circ}$, 
where also the bulk of the quasar fields
are found. The remaining two lie at $|b| \approx 30^{\circ}$, and 
were obtained in order to 
get a handle on the counts here since two fields in our radio galaxy sample
(to be published in a future paper) lie at lower latitudes.

The counts in the two $|b| \approx 30^{\circ}$ fields have a mean of 
16.94$\pm$2.3 arcmin$^{-2}$, whereas the fields between 45 and 60$^{\circ}$ 
have a mean of 
11.53$\pm$1.6 arcmin$^{-2}$, so the background counts appear to 
increase towards lower latitudes. Presumably this is due to faint, red stars 
as the background counts in $R$ do not seem to vary as much with latitude.
It is seen that the 
bulk of the quasar and control fields lie at latitudes between 
40 and 60$^{\circ}$. Nevertheless, in addition to using the average 
background counts, as we did in Paper~I, we also here make a 
correction for the variation in the counts with latitude. 

In $I$-band, a least square fit to the eight data points 
weighted with 
the error bars gives a relation for the background counts of 
$N\left(I\right)=-0.32|b|+26.6$.
This is shown by the solid line in 
Fig.~\ref{fig:fig2}, and we call this Model~1.
A straight line through the mean values of the two low latitude 
and the six high latitude fields 
gives a less steep relation of $N\left(I\right)=-0.25|b|+24.5$,
as shown by the dotted line. This relation is named Model~2.

As mentioned above, the $R$-band background counts show a weaker 
dependence on galactic
latitude. The corresponding Model~1 for the $R$ counts is 
$N\left(R\right)=-0.15|b|+18.6$, and by drawing a line through the
mean of the seven fields at 40--60$^{\circ}$ and 
the mean of the two $|b|\approx 30^{\circ}$ fields, we get that 
Model~2 for the $R$-band 
counts is $N\left(R\right)=-0.17|b| + 19.4$ galaxies per arcmin$^{2}$. 
Outside the ranges of the sloped lines in Fig.~\ref{fig:fig2},
we have used the constant values corresponding to the endpoints of the
relations. 

\begin{table*}
\begin{minipage}{12.7cm}

\caption{Various relations used for the galaxy background counts. 
In Model~1 and 2 the counts are dependent on the absolute value of 
the galactic latitude $|b|$, whereas Model~3 is just the average counts. 
The error in the average given in the last column is the standard deviation 
in the mean, and the other errors were calculated as 1.3$\sqrt{N}$.}

\begin{tabular}{llllll}
& \multicolumn{2}{c}{Model 1} & \multicolumn{2}{c}{Model 2} & Model 3 \\
Filter & Range ($^{\circ}$)     &   $N$ (arcmin$^{-2}$)     &   Range ($^{\circ}$) &  $N$ (arcmin$^{-2}$) & $N$ (arcmin$^{-2}$) \\
 & &  &  &   \\
$I$ & 30.0$\leq |b| \leq$57.1 & $-$0.32$|b|+$26.6   & 30.1$\leq |b| \leq$51.6  & $-$0.25$|b|+$24.5 & 12.36$\pm$4.38 \\
$I$ & $|b|<$30.0              & 17.07$\pm$5.37                & $|b|<$30.1         &  16.94$\pm$5.35   & 12.36$\pm$4.38 \\
$I$ & $|b|>$57.1              & 8.50$\pm$3.79                 & $|b|>$51.6         &  11.53$\pm$4.41   & 12.36$\pm$4.38 \\
    &                   &                                     &      \\
$R$ & 29.2$\leq |b| \leq$57.2 & $-$0.15$|b|$+18.6    & 29.7$\leq |b| \leq$48.5  &  $-$0.17$|b|$+19.4& 12.51$\pm$2.69 \\
$R$ & $|b|<$29.2              & 14.07$\pm$4.88                & $|b|<$29.7         & 14.45$\pm$4.94 & 12.51$\pm$2.69 \\
$R$ & $|b|>$57.2              & 9.74$\pm$4.06                 & $|b|>$48.5         & 11.30$\pm$4.37 & 12.51$\pm$2.69 \\
    &                   &                                     &      \\
$V$ & 30.0$\leq |b| \leq$57.4 & $-$0.15$|b|$+18.23 &  & & 10.19$\pm$6.88 \\   
$V$ & $|b|<$30.0              & 13.60$\pm$4.79     &  & & 10.19$\pm$6.88 \\ 
$V$ & $|b|>$57.4              & 9.40$\pm$4.0       &  & & 10.19$\pm$6.88 \\ 
\end{tabular}
\label{table:table3}
\end{minipage}
\end{table*}

We have chosen to let the background counts follow these relations 
when correcting for background galaxies in the quasar fields, but 
we have also used the average of all control fields, named Model~3.
Details about these models can be found in 
Table~\ref{table:table3}. We have also included the $V$-band control fields 
in this table since these were used to find the excess in five
quasar fields that were also imaged in $R$. Since there are only two
$V$-band control fields (at $|b|=57\fdg4$ and 46\fdg7), we have used 
as Model~1, a relation that has the same slope as the $R$ counts, but is 
normalized to the mean $V$-band counts.

\section{Analysis}

\label{section:s4}

To quantify the excess counts in the quasar fields we have used 
the amplitude, $B_{\rm gq}$, of the spatial galaxy--quasar cross-correlation 
function, $\xi\left(r\right)=B_{\rm gq} r^{-\gamma}$,
where $\gamma=1.77$. The amplitude is evaluated at a fixed radius of 0.5 Mpc 
at the quasar redshift, corresponding to $\approx 1$ arcmin at $z=0.7$ and has units
of Mpc$^{1.77}$.
Longair \& Seldner \shortcite{ls79} showed how it can be used to estimate the
clustering about any point in the Universe using galaxy counts, and it has subsequently 
been widely used to quantify galaxy environment around different types of AGN 
(e.g. Yee \& Green 1987; Yates, Miller \& Peacock 1989; 
Ellingson et al.\ 1991a; 
Smith, O'Dea \& Baum 1995; Wurtz et al.\ 1997; DeRobertis, Yee \& Hayhoe 1998; 
McLure \& Dunlop 2000).
In Paper~I we used it to quantify the environments around RLQs, and
here we apply the same analysis to the RQQs. 
We therefore refer to Paper~I for a more detailed description of the analysis,
and give only a brief summary here 
(see also Longair \& Seldner 1979; Yee \& Green 1987).

First, the amplitude, $A_{\rm gq}$, of the {\em angular} cross-correlation 
function is found. This is 
easily obtainable from the data since it is directly proportional 
to the number of 
excess galaxies within the 0.5 Mpc radius centered on the quasar.
Thereafter, $B_{\rm gq}$ is obtained from the following expression
\[
B_{\rm gq}=\frac{{\cal{N}_{\rm g}}A_{\rm gq}}{\Phi\left(m_{\rm lim},z\right)I_{\gamma}}d_{\theta}^{\gamma-3},
\]
\noindent
where $\cal{N}_{\rm g}$ is the average surface density of background galaxies, 
$d_{\theta}$ is the angular diameter 
distance to the quasar and $I_{\gamma}=3.78$ is an integration constant. 

In the denominator, $\Phi\left(m_{\rm lim},z\right)$ is the luminosity function
integrated down to the completeness limit of the data, thus giving the 
number of galaxies per unit volume at the quasar redshift seen
in the flux-limited data.
We have used a Schechter function to evaluate the integral, with characteristic magnitude
$M^{*}_{AB}\left(B\right)=-20.83$ and slope $\alpha=-0.89$ based on Lilly et al's 
\shortcite{lilly95} estimation of the CFRS (Canada-France Redshift Survey)
luminosity function in the $0.5\leq z \leq 0.75$ redshift interval. 
Since the integrated Schechter function is proportional to the characteristic
density, $\phi^{*}$, the $B_{\rm gq}$ estimator becomes inversely proportional to it,
and therefore very sensitive to its value. 
Yee \& Green \shortcite{yg87} and Yee \& L{\'o}pez-Cruz \shortcite{ylc99} have advocated
the importance of choosing $\phi^{*}$ in such a way that it is consistent with
the data. In Paper~I, we explain how we used the CFRS luminosity function to 
construct galaxy counts that were fitted
to the observed counts in the control fields. The fitting gave 
$\phi^{*}_{V}=0.0065$, $\phi^{*}_{R}=0.0072$ and 
$\phi^{*}_{I}=0.0052$ Mpc$^{-3}$, and we use these numbers in our computation of
$\Phi\left(m_{\rm lim},z\right)$.

\begin{figure}
\psfig{file=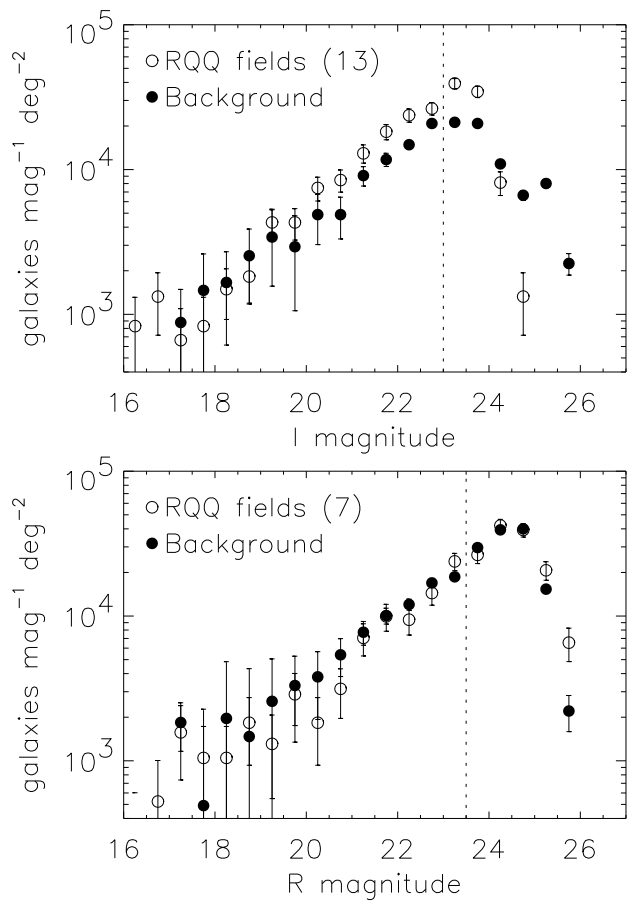}
\caption{The open circles show the average galaxy counts within the 0.5 Mpc 
circle in the quasar fields, and the filled circles show the counts in the
control fields.
The vertical dotted lines mark the completeness limits, $I=23.0$ and $R=23.5$.}
\label{fig:fig3}
\end{figure}

Yee \& L{\'o}pez-Cruz \shortcite{ylc99} have examined the robustness of 
$B_{\rm gc}$ (index `gc' means galaxy--cluster centre) on a sample of 
47 low-$z$ Abell clusters, and find that it is a robust estimator of galaxy
richness in clusters. They apply a number of 
tests, and find that $B_{\rm gc}$ is fairly robust provided the 
normalization of the luminosity function reproduces the observed galaxy counts.
Under this condition, varying e.g.\ $M^{*}$ causes variations in $B_{\rm gc}$ of 
$\approx 10$ per cent. Also counting down to different absolute magnitude 
limits produces systematic variations of only a few per cent, but the best and most 
reliable results are obtained if one counts down to 1--2 magnitudes fainter than 
$M^{*}$. In the $I$-band images we have reached approximately 1.7 mag past $M^{*}$, 
and in the $R$-band
data, about 2 mag (see last column of Table~\ref{table:table4}).


\section{Galaxy excess in quasar fields}

\label{section:s5}

We counted the number of galaxies within the 0.5 Mpc
radius in the quasar fields and averaged the counts for the $R$ and $I$-band data, 
i.e.\ for the seven $z<0.67$ and the 13 $z \ge 0.67$ quasar fields. The average counts are
shown in Fig.~\ref{fig:fig3}, where we also have plotted the average 
galaxy counts in the control images. 
The error bars of the background galaxy count were calculated as 
1.3$\sqrt{N}$. There is a clear excess of faint galaxies at $I>20$, but the 
galaxy counts in the $R$-band fields show no average excess above the background. 

For each quasar field, the net excess of galaxies within the 0.5 Mpc radius, 
$N_{\rm net}$, was calculated 
by subtracting the background counts according to Model~1--Model~3 from the counts 
in the quasar fields. Thereafter, the amplitudes $A_{\rm gq}$ and
$B_{\rm gq}$ were obtained. The results are shown in Table~\ref{table:table4},
where we, for each model of the background counts, show the net excess, 
$N_{\rm net}$, the significance, $\sigma$, of the excess, and the
clustering amplitude, $B_{\rm gq}$. 

Since the background counts were scaled to the area of the 0.5 Mpc circle, 
$N_{\rm net}$ is not an integer. The significance, $\sigma$, was calculated
as $N_{\rm net}/1.3\sqrt{N_{\rm b}}$, and the 
uncertainty in $B_{\rm gq}$ as
\[
\frac{\Delta B_{\rm gq}}{B_{\rm gq}}=\frac{\left(N_{\rm net}+1.3^{2}N_{\rm b}\right)^{1/2}}{N_{\rm net}}.
\]
\noindent
\cite{ylc99}.

In the last column of Table~\ref{table:table4}, we show approximately how many magnitudes
fainter than $M^{*}$ an Sa galaxy with apparent magnitude $m=m_{\rm lim}$ is.
This demonstrates that we
are close to the recommended $M^{*}+2$ in all the fields so that a reliable estimate of
$B_{\rm gq}$ is obtained.
By transforming $M^{*}_{AB}\left(B\right)=-20.83$ to $R$ and $I$, 
assuming $B_{AB}=B-0.17$ \cite{oke72} and using colours of a zero redshift 
Sa galaxy model, $B-V=0.74$, $V-R=0.68$ and $R-I=0.57$ \cite{grv88}, we found that 
$M^{*}_{R}=-22.08$ and $M^{*}_{I}=-22.65$.

\begin{table*}
\begin{minipage}{16.8cm}

\caption{The results of the analysis of galaxy counts in the
RQQ fields. For each of the three models for the background counts
(see Table~\protect\ref{table:table3}), the net excess, $N_{\rm net}$, the 
significance of the excess, $\sigma$, and the clustering amplitude, 
$B_{\rm gq}$, are listed. 
The absolute $B$ magnitude of the quasar is $M_{B}$, and $\Delta M$
in the last column shows how many magnitudes into the luminosity 
function the completeness limit extends, i.e. $M^{*}+\Delta M$.
All numbers are based on galaxy counts in either $R$ or $I$ depending 
on whether the quasar redshift is $<0.67$ or $\ge 0.67$.}

\begin{tabular}{llllllllllll}
\multicolumn{1}{c}{ } & \multicolumn{1}{c}{ } & \multicolumn{3}{c}{Model 1} & \multicolumn{3}{c}{Model 2} & \multicolumn{3}{c}{Model 3} & \multicolumn{1}{c}{$\Delta M$} \\
     Quasar             &         $M_{B}$             & $N_{\rm net}$ & $\sigma$ & $B_{\rm gq}$ & $N_{\rm net}$ & $\sigma$ & $B_{\rm gq}$ & $N_{\rm net}$ & $\sigma$ & $B_{\rm gq}$ \\
                                  &           &      &  &  &  &   & & &  & &  \\ 
LBQS 0007$-$0003  &   $-$24.58 & 3.9    & 0.6    & 84$\pm$155 & 
$-$6.4  & $-$0.8 & $-$137$\pm$165 & 
$-$6.9 & $-$0.8 & $-$146$\pm$166& 1.9\\

LBQS 0020$-$0300  & $-$24.94 & $-$4.4 & $-$0.6 & $-$68$\pm$119 & 
$-$10.4 & $-$1.2 & $-$161$\pm$123  & 
$-$14.8 & $-$1.7 & $-$230$\pm$126 & 1.9\\

BFSP 12344$-$0021  & $-$23.53 & 21.0  & 3.1 & 529$\pm$208 & 
11.1  & 1.4 & 278$\pm$218  & 
10.7 & 1.3 & 268$\pm$219 & 1.7 \\

BFSP 12355$+$0107 & $-$22.90 & 16.4 & 2.4 & 371$\pm$182 & 
6.2     & 0.8  & 140$\pm$192 & 
5.8 & 0.7 & 131$\pm$192 &  1.8\\

BFSP 12364$-$0053 & $-$22.18 & 24.5 & 3.5 & 565$\pm$197 &
14.3   & 1.8  & 331$\pm$206 & 
13.9 & 1.7 & 321$\pm$206 &  1.8\\

BQS 1333$+$176    &  $-$25.96$^{\star}$\footnotetext{$^{ \star}$Found from Gunn $r$ magnitude as observed by Yee et al.\ (1986)} & 1.4 & 0.2 & 20$\pm$116 & 
$-$4.8 & $-$0.6  & $-$68$\pm$119 & 
$-$9.4 & $-$1.0 & $-$133$\pm$122 & 2.1\\

BFSP 15152$+$0244 &  $-$22.47  & 8.3 & 1.0 & 144$\pm$156 & 
8.7    & 1.0 & 151$\pm$156  & 
4.8 & 0.5 & 83$\pm$158 & 1.8 \\

BFSP 15196$+$0220 &  $-$23.55  & 17.6 & 2.2 & 428$\pm$222 & 
13.9   & 1.6 & 337$\pm$226  & 
18.3 & 2.3 & 443$\pm$222 &  1.7\\

BFSP 15199$+$0247 & $-$23.56  & 29.6 & 3.7 & 777$\pm$255 & 
25.8   & 3.1 & 678$\pm$259  & 
29.9 & 3.7 & 785$\pm$255 &  1.6\\

BFSP 15202$+$0236 &  $-$22.60  & 26.4 & 3.3 & 689$\pm$250 & 
22.7  & 2.7 & 592$\pm$253 & 
26.9 & 3.4 & 703$\pm$250 &  1.6\\

BQS  1538$+$477   &  $-$27.29$^{\dagger}$\footnotetext{$^{\dagger}$Computed from $B$ found by Schmidt \& Green (1983)}  & 48.0 & 6.4 & 1276$\pm$271 & 
43.1 & 5.4 & 1147$\pm$275 & 
43.0 & 5.4 & 1144$\pm$275 &  1.6 \\

BFSP 22011$-$1857 & $-$23.46 & 8.5 & 1.0 & 152$\pm$157 & 
7.3   & 0.9 & 131$\pm$158 & 
3.1 & 0.4 & 55$\pm$161 &  1.7\\

LBQS 2235$+$0054  &  $-$22.29  & 13.6 & 1.5 & 176$\pm$125 & 
14.1   & 1.6 & 182$\pm$125 & 
9.8 & 1.1 & 127$\pm$127 &  2.2\\

LBQS 2238$+$0133  &  $-$25.48  & 10.3 & 1.3 & 229$\pm$194 & 
6.2     & 0.7  & 138$\pm$197 & 
9.6 & 1.2 & 215$\pm$194 &  1.8\\

LBQS 2239$-$0055  & $-$24.66  & 22.3 & 2.8 & 450$\pm$187 & 
17.6   & 2.1  & 356$\pm$191 & 
19.5 & 2.4 & 394$\pm$190 &  2.0\\

LBQS 2245$-$0055  & $-$26.07  & $-$1.8 & $-$0.2 & $-$52$\pm$219 & 
$-$6.3  & $-$1.5 &  $-$185$\pm$225 & 
$-$5.4 & $-$0.7 & $-$158$\pm$224 & 1.5\\

LBQS 2348$+$0210  &  $-$24.40$^{\dagger\dagger}$\footnotetext{$^{\dagger\dagger}$Computed from $B_{J}$ given by Hewett et al. (1995)} & 4.2 & 0.5 & 50$\pm$105 & 
$-$1.2 & $-$0.1  &   $-$14$\pm$107 & 
$-$6.1 & $-$0.7 & $-$73$\pm$109 &  2.4\\

LBQS 2348$+$0148  &  $-$24.42 & 6.9 & 1.0 & 172$\pm$183 & 
$-$3.1  & $-$0.4 & $-$76$\pm$194 & 
$-$3.5 & $-$0.4 & $-$86$\pm$194 & 1.7\\

LBQS 2350$-$0012  & $-$25.30 & 1.7 & 0.2 & 25$\pm$118 & 
$-$4.4  & $-$0.5 & $-$64$\pm$122 & 
$-$9.0 & $-$1.0 & $-$130$\pm$124 & 2.0\\

LBQS 2353$-$0153  &  $-$26.37 & 35.3 & 5.0 & 699$\pm$183 & 
24.7  & 3.0    & 489$\pm$191 & 
24.2 & 2.9 & 480$\pm$191 & 2.0\\
\end{tabular}
\label{table:table4}
\end{minipage}
\end{table*}

The mean clustering amplitude using Model~1 for the background counts is 
336$\pm$77 Mpc$^{1.77}$,
whereas using Model~2 and Model~3 gives lower mean amplitudes of 212$\pm$74 and 210$\pm$82
Mpc$^{1.77}$. The correction in the background counts to account for 
galactic latitude is therefore
more drastic in Model~1, giving a higher mean $B_{\rm gq}$, but
the three mean values are nevertheless within the errors of each other. 
(The errors we give for the mean $B_{\rm gq}$'s are
standard deviations in the mean.)
Also, for each field, the three $B_{\rm gq}$'s given in Table~\ref{table:table4} are 
consistent with each other, and their variation should give some indication of 
the real errors that are involved in this analysis. 

The mean clustering amplitude in the quasar fields, regardless of which model
is used for the background counts, is higher than the local galaxy--galaxy 
correlation amplitude of $\approx60$--70 Mpc$^{1.77}$ 
(Davis  \& Peebles 1983; Loveday et al.\ 1995; Guzzo et al.\ 1997),
so on average, we find that the RQQs 
are not located in field-like environments, but rather seem to 
prefer groups or poorer clusters.

Using a $\chi^{2}$ test, we compared the distribution of the 
significances of the excess to that of a normal distribution 
(which is what we expect if there are no excess galaxies around the 
quasars), and find a probability of 9$\times$10$^{-20}$ (Model~1)
that the distribution is equal to that of a normal distribution.
For Model~2 and 3, we get probabilities of 
9$\times$10$^{-9}$ and 1$\times$10$^{-10}$.
Excluding the seven radio-detected quasars, 
we get probabilities (for Model~1 to 3) of 1.0$\times$10$^{-10}$, 
1.1$\times$10$^{-3}$ and 4.4$\times$10$^{-5}$
that the distribution of $\sigma$'s is equal to a normal distribution.
Furthermore, excluding the two BQS quasars, the probabilities 
become 5$\times$10$^{-10}$, 5$\times$10$^{-4}$ and 7$\times$10$^{-6}$.
These very low probabilities agree with the visual impression 
from Fig.~\ref{fig:fig8}, where the $B_{\rm gq}$'s of both RLQs and RQQs
are plotted as a function of redshift and $M_{B}$.
The low probabilities are a strong result, and can imply either 
an average density enhancement or a variance in the environments 
greater than expected, and most likely, both are present in the data.

This adds more weight to the result that RQQs are on average not located
in field-like environments. However, the environmental richness varies between
individual fields; some fields have no significant excess, whereas other
appear to be very rich in galaxies.
There are five fields with a 
$> 2$$\sigma$ excess irrespective of how the background
is calculated. The richest field among these is the field around BQS 1538$+$447 with a 
clustering amplitude in the range 1100--1200 Mpc$^{1.77}$ and a galaxy excess of $\approx 40$--50.
In Paper~I we argued that an amplitude of $\approx740$ Mpc$^{1.77}$ corresponds roughly 
to Abell richness class $\ga$ 1, so the cluster candidate around BQS 1538$+$447 may even qualify 
as an Abell class 2 cluster. 
Four other fields have galaxy excesses of 20--30 galaxies and amplitudes 
$\ga$ 500 Mpc$^{1.77}$
and are probable Abell class 1 clusters. 
BFSP 15199$+$0247 and 15202$+$0236 have amplitudes in the range 700--800 Mpc$^{1.77}$,
whereas LBQS 2239$-$0055 and 2353$-$0153 have somewhat smaller 
amplitudes of 400--500 Mpc$^{1.77}$.

As Fig.~\ref{fig:fig3} shows, the $R$-band fields have a galaxy 
surface density close to that in the control fields, and the mean
clustering amplitude for these, 71$\pm$34 Mpc$^{1.77}$, is consistent with that of
galaxies in the local universe (23$\pm$50 and $-$43$\pm$50 Mpc$^{1.77}$ for 
Model~2 and 3). 
For the 13 $I$-band fields the mean clustering amplitude is 
478$\pm$96 Mpc$^{1.77}$
(314$\pm$101 and 346$\pm$106 Mpc$^{1.77}$ for Model~2 and 3, respectively).
It might therefore seem that there is an apparent trend with redshift since the $R$-band 
fields represent the $z<0.67$ quasars, but the current sample of 20 objects is too small and 
span a too narrow redshift range to test this properly.

In principle, we are interested in the full distribution function for the clustering
amplitude, e.g.\ when the data are compared with models.
The distribution will 
be positively skewed with a long tail to the right since we do not observe fields with
negative $B_{\rm gq}$'s comparable in absolute value to the most positive $B_{\rm gq}$ values. 
For a positively skewed distribution, the median will always be smaller than the 
mean, and for our sample the median $B_{\rm gq}$ is 131 Mpc$^{1.77}$ (Model~3).
A model for formation of RQQs should be able to reproduce both the mean amplitude and
the intrinsic scatter in $B_{\rm gq}$. Thus the intrinsic scatter is equally important 
as the mean amplitude. The scatter (standard deviation) of the $B_{\rm gq}$ distribution  
is $\sigma_{B_{\rm gq}}=365\pm58$ Mpc$^{1.77}$ (Model~3), and this includes both 
intrinsic scatter and measuring uncertainty. 
We have used a measuring uncertainty of $\Delta B_{\rm gq} \approx 185\pm48$ Mpc$^{1.77}$, and 
estimated the intrinsic scatter, $\sigma_{\rm int}$, assuming that 
$\sigma_{\rm int}^{2} = \sigma_{B_{\rm gq}}^{2} + \Delta B_{\rm gq}^{2}$, giving
$\sigma_{\rm int}=315\pm73$ Mpc$^{1.77}$. As expected, the intrinsic scatter is large, but 
difficult to estimate due to the large measuring uncertainty in $B_{\rm gq}$. 
In Table~\ref{table:table6} we list values of 
$\sigma_{\rm int}$, mean and median $B_{\rm gq}$'s 
that we get by using the different models for the background
galaxy counts.

Since 12 of the fields were imaged in two filters, we also computed  
$B_{\rm gq}$ in the fields taken in the shortest wavelength filter,
and list the results in Table~\ref{table:table5}.
The mean amplitudes are 220$\pm$92 and 120$\pm$93 Mpc$^{1.77}$ for Model~1 and 
Model~3, respectively (Model~2 gives values intermediate between these two).
For comparison, the mean amplitudes for the same 12 fields obtained in
the reddest filters are 367$\pm$120 and 
276$\pm$125 Mpc$^{1.77}$. So, the mean amplitudes are lower, but still within the
errors of each other, and on average, the excess is detected in both filters. 
There are two fields which have significantly higher $B_{\rm gq}$ in $I$ than in $R$, and 
these are BQS 1538$+$447 and LBQS 2353$-$0153, with $B_{\rm gq}$ approximately
a factor of two higher in $I$ than in $R$, suggestive of a fair amount of 
red galaxies in these fields. 
This is what we expect if there are many red, early-type
galaxies with strong 4000 {\AA} breaks at the quasar redshift.

\begin{table*}
\begin{minipage}{11.8cm}

\caption{The results of the analysis of the quasar fields that 
were imaged in two filters, here in the shortest wavelength filter, i.e.\ in $V$ for the 
$z<0.67$ quasars and in $R$ for the $z \ge 0.67$ quasars. 
The completeness limit in $V$ is 24.0.}

\begin{tabular}{llllllll}
\multicolumn{1}{c}{ } & \multicolumn{1}{c}{ } & \multicolumn{3}{c}{Model 1} & \multicolumn{3}{c}{Model 3} \\

Quasar    & Filter & $N_{\rm net}$ & $\sigma$ & $B_{\rm gq}$ & $N_{\rm net}$ & $\sigma$ & $B_{\rm gq}$ \\
          &        &               &          &              \\
LBQS 0020$-$0300  & $V$ & $-$13.0 & $-$1.7  & $-$341$\pm$181  & $-$16.1 & $-$2.0 & $-$420$\pm$185 \\
BFSP 15199$+$0247 & $R$ & 22.0    & 2.7     & 594$\pm$251   & 19.3 & 2.3 & 521$\pm$254 \\
BFSP 15202$+$0236 & $R$ & 16.9    & 2.1     & 449$\pm$240    & 14.3 & 1.7 & 380$\pm$243 \\
BQS  1538$+$477   & $R$ & 27.3    & 3.7     & 745$\pm$246   & 18.4 & 2.2 & 503$\pm$255 \\
BFSP 22011$-$1857 & $V$ & 14.6    & 1.8     & 464$\pm$283   & 15.4 & 1.9 & 490$\pm$282 \\
LBQS 2235$+$0054  & $V$ & 10.0     & 1.2     & 201$\pm$187 & 13.2 & 1.6 & 265$\pm$184 \\
LBQS 2238$+$0133  & $R$ & 1.2     & 0.1     & 25$\pm$175      & $-$2.1 & $-$0.3 & $-$44$\pm$178 \\
LBQS 2239$-$0055  & $R$ & 17.2    & 2.3     & 322$\pm$162  & 7.7 & 0.9 & 144$\pm$169 \\
LBQS 2245$-$0055  & $R$ & $-$0.2  & $-$0.0  & $-$7$\pm$229  & $-$8.9 & $-$1.1 & $-$283$\pm$242 \\
LBQS 2348$+$0210  & $V$ & 8.8     & 1.1     & 157$\pm$157     & 5.6 & 0.7 & 99$\pm$159 \\
LBQS 2350$-$0012  & $V$ & $-$7.9 & $-$1.0  & $-$187$\pm$175 & $-$11.0 & $-$1.3 & $-$261$\pm$178 \\
LBQS 2353$-$0153  & $R$ & 11.9    & 1.6     & 218$\pm$152   & 2.4 & 0.3 & 44$\pm$159 \\
\end{tabular}
\end{minipage}
\label{table:table5}
\end{table*}

Although the overall trend is that $B_{\rm gq}$ is larger in the reddest filters,
there are 3--4 fields where the opposite is seen.
This can, apart from in one case, 
probably be attributed to measurement errors since these are fields
with no significant excess. The exception is 
BFSP 22011$-$1857, with a $\sim 2\sigma$ excess 
in $V$, but with no excess in $R$.
The clustering amplitude deduced from the $R$ image lies in the
range 50--150, whereas it is $\approx 470$ Mpc$^{1.77}$ in $V$-band.
This field was carefully checked in order to verify that there are 
more galaxies 
in the $V$ image than in the $R$ image. There are two possible explanations for this,
one is that there is a foreground concentration of bluer galaxies toward this quasar, 
the other is that it might be a cluster with a blue galaxy population
at the quasar redshift.


\subsection{Properties of the excess galaxies }

\label{section:s51}

Elliptical and early-type spiral galaxies in clusters are known to form
a tight sequence in the color-magnitude diagram, the so-called `red sequence'.
It is the strongly evolved 4000 {\AA} breaks in these galaxies
that make their colour red, since metals in the cool atmospheres of giant stars
absorb much of the radiation blueward of 4000 {\AA}. 

In the conventional models, 
elliptical galaxies are believed to form during a monolithic collapse 
at $z>2$ and evolve passively after an initial
starburst (e.g.\ Larson 1975; Arimoto \& Yoshii 1987). 
An alternative model that has received much attention lately is 
the hierarchical formation scenario, where
elliptical galaxies are formed by mergers of disk systems 
(Toomre \& Toomre 1972).
It has been shown that this scenario is also consistent with
the presence of the red sequence in galaxy clusters \cite{kc98}. 

\begin{figure}
\psfig{file=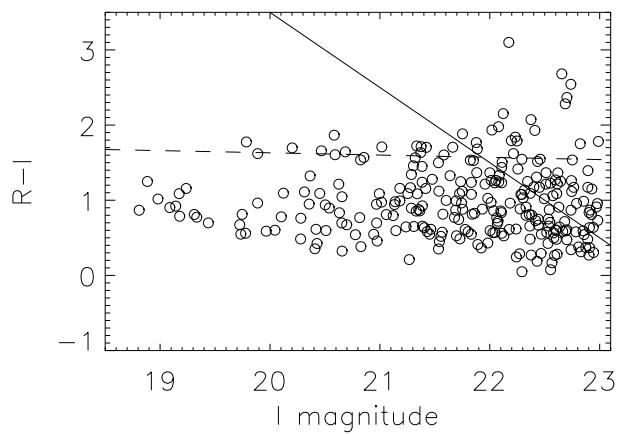}
\caption{Colour-magnitude diagram of galaxies in the four richest 
quasar fields, showing a hint of a red sequence at $R-I\approx 1.5$--1.7. 
Possible stars at $I<21$ are excluded. The dashed line shows a least 
square fit to the galaxies at $19 < I \leq 21.5$ and $1.3 < R-I < 2.0$ 
with a slope of $-$0.07$\pm$0.03. The solid line marks the completeness limit in 
$R$.}
\label{fig:fig4}
\end{figure}

Here, we have picked the four richest quasar fields that were 
imaged in two filters,
and examined the colours of the galaxies. The four fields are BFSP 15199$+$0247, 
BFSP 15202$+$0236, BQS 1538$+$447 and LBQS 2353$-$0153, with 
$B_{\rm gq} \ga 500$ Mpc$^{1.77}$
and a $\ga 3 \sigma$ excess. The mean redshift for these quasars is 
$\left<z\right>=0.74\pm0.02$.
Fig.~\ref{fig:fig4} shows a colour-magnitude diagram of galaxies in these fields, where we have excluded stars at $I<21$ (as explained later in this
section). 

There is a hint of a red sequence at $R-I \approx 1.5$--1.7 in this figure, 
i.e.\ tentative evidence that these fields likely contain galaxy 
clusters at $z\approx 0.7$--0.8.  
Very few galaxy clusters are know at $z \ga 0.7$, but judging from other 
work, the red sequence is expected to occur at $1.3 \la R-I \la 2.0$.
(Luppino \& Kaiser 1997; Clowe et al.\ 1998, Lubin et al.\ 2000). 
Also spectral synthesis models predict that this is the colour 
expected for early-type galaxies at redshifts 0.7--0.8 
(e.g.\ Fukugita, Shimasaku \& Ichikawa 1995).

The red sequence is known to steepen toward brighter magnitudes, and 
this can possibly also be seen in Fig.~\ref{fig:fig4}. The steepening toward 
brighter magnitudes is believed to be caused by the mass-metallicity relation in 
the galaxies, the most luminous and massive galaxies having a higher metal 
content and therefore also stronger 4000 {\AA} breaks.
The slope of the red sequence in clusters is also expected to flatten
toward higher redshifts. As the formation epoch of a galaxy is approached, the 
stellar populations in the massive ellipticals become younger and bluer relative
to the low-mass systems, and by measuring the slope of the red sequence as a function
of redshift, one can constrain the formation epoch of the galaxies.
Gladders et al.\ \shortcite{gladders98} have studied clusters in 
the redshift range
$0 < z < 0.75$, and find that there is little evidence for a change in the slope 
prior to $z=0.75$, thereby concluding that the formation epoch occurs at 
$z > 2$. For two clusters at $z=0.756$, Gladders et al.\ find a slope of 
$-$0.0566$\pm$0.0066, and a least square fit to the galaxies at $19 \leq I \leq 21.5$
and $1.3 < R-I < 2.0$ in Fig.~\ref{fig:fig4} gives a slope of $\approx -$0.07$\pm$0.03,
consistent with Gladders et al. 

Ferreras \& Silk \shortcite{fs00} also show that the red sequence in clusters
can by used to discriminate between the two formation scenarios, monolithic vs
hierarchical, particularly at the bright end.
The two formation scenarios predict 
significantly different bright-end slopes at high redshifts, and a number of 
$z > 0.7$--0.8 clusters must be observed to test this. 
Since few clusters are known at these redshifts, AGN-selected clusters are therefore 
useful targets for such studies.    

\begin{figure}
\psfig{file=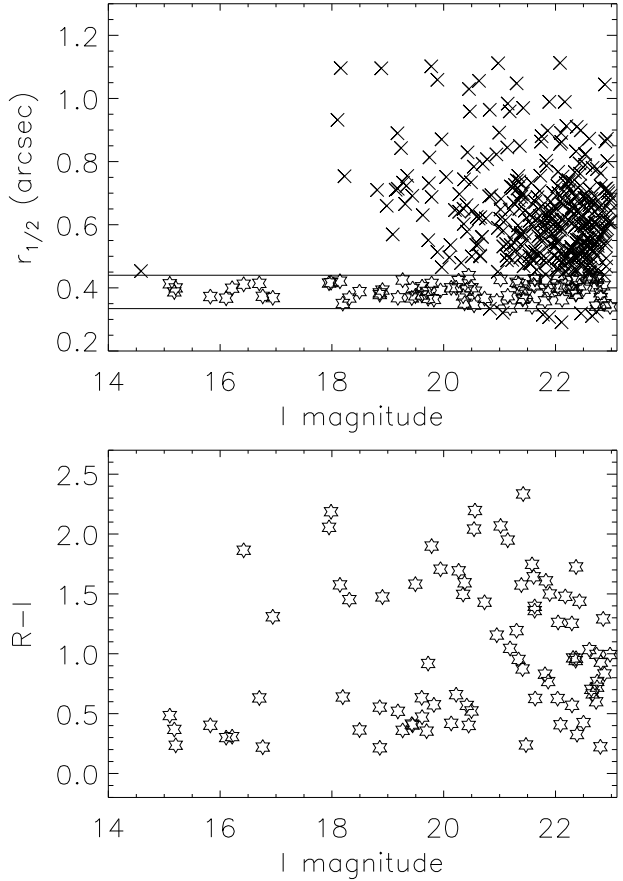}
\caption{The upper plot shows the half-light radius of the objects 
in the four richest quasar fields as a function of magnitude. 
The stellar locus is clearly seen as the objects between the two 
horizontal lines (i.e.\ at seeing FWHM $\approx 0.8$ arcsec).
The lower plot shows a colour-magnitude diagram of the objects on the 
stellar locus.} 
\label{fig:fig5}
\end{figure}

Can the tentative red sequence at $R-I \approx 1.5$--1.7 be caused 
by stars? This is 
probably not the case, since we do not expect many stars in these high latitude 
fields (these four quasar fields lie at latitudes 
46--61$^{\circ}$).
Nevertheless, the $I$-band counts in the control fields seem to indicate that red stars
are present in the fields at some level. 
To investigate this further, we 
made an attempt to identify stars in these images by their 
half-light radii, as shown in Fig.~\ref{fig:fig5}. 

This was done using the {\sc imcat} software by
N. Kaiser, developed to measure the distortion of faint galaxies 
induced by gravitational potentials of galaxy clusters. The software
automatically measures the half-light radii of galaxies in the images,
a feature that is not included in {\sc focas}. However, {\sc imcat} 
uses a different
detection algorithm than {\sc focas}. In {\sc focas}, objects are detected by
locating connected regions above a certain threshold, whereas {\sc imcat} smooths
the images with gaussian filters using a range of different radii 
(see Kaiser, Squires \& Broadhurst (1995) for details).

In the upper plot in Fig.~\ref{fig:fig5}, the stellar locus is seen along a half-light 
radius of $\approx 0.4$ arcsec, corresponding to a seeing FWHM of $\approx 0.8$ arcsec,
and stars can clearly be separated from galaxies down to $I\approx 21$.
We selected objects with half-light radii between the two horizontal, solid lines
(corresponding to half-light radii of 1.9 and 2.5 pix, or 0.33 and 0.44 arcsec)
and plotted them in a colour-magnitude diagram as shown in the lower plot. 
The colour-magnitude distribution of stars is expected to be bimodal,
with disk stars at the blue end and spheroid stars at the red end
\cite{bs81}. The bimodal distribution is not clearly seen here, 
presumably due to the small number of stars, but there seems to be 
a gap at $0.8 < R-I < 1.3$. Bahcall \& Soneira find that
the stars in the Galactic spheroid have typical colours of 
$R-I \approx 1.6$--2.2, and that the bluer disk population has 
$R-I \approx 0.1$--0.8, which appears to agree roughly with what we see here.

At $19 < I \leq 23$ and at colours $1.3 < R-I < 2.0$, where we suspect the
red sequence to be, 18 out of 90 objects have half-light radii corresponding
to the stellar locus in Fig.~\ref{fig:fig5}, i.e.\ four--five
stars per field, so the red sequence cannot be caused exclusively by stars.
Since {\sc imcat} appears to have done a very good job at identifying
the stars, we omitted the objects on the stellar locus at $I < 21$ 
in the colour-magnitude diagram in Fig.~\ref{fig:fig4}.
The $R-I$ colours were evaluated using aperture 
magnitudes measured by {\sc imcat} in a 1.5 arcsec aperture, corresponding
to approximately twice the FWHM of the seeing disk.

\begin{figure}
\psfig{file=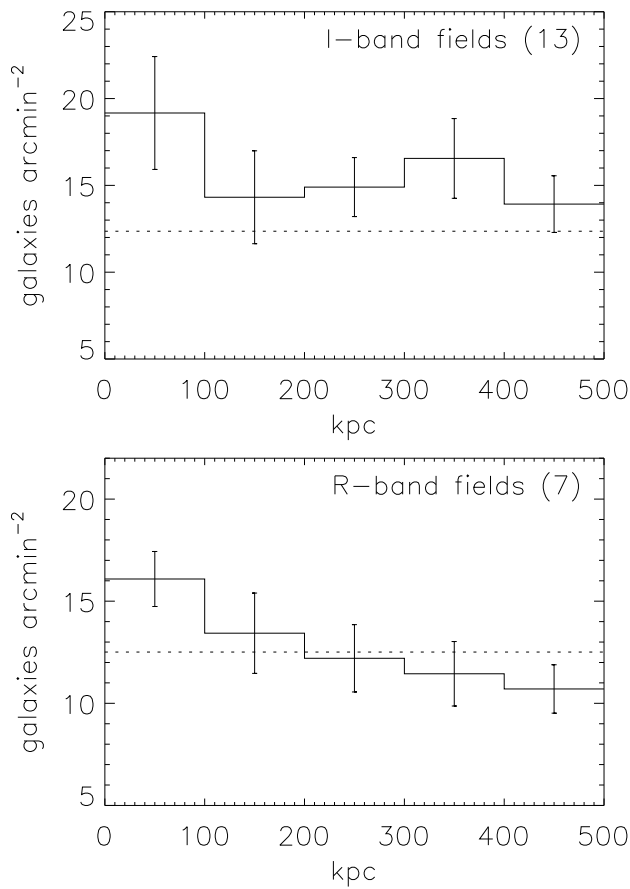}
\caption{Average radial surface density profiles for the quasar fields.
The quasars are located in the inner bins, and the error bars show the
standard deviation in the mean. 
The dotted horizontal lines correspond to the average 
background count.}
\label{fig:fig6}
\end{figure}

We also examined the radial distribution of galaxies in the quasar 
fields as shown
in Fig.~\ref{fig:fig6}. Here, we have taken the average of all fields, and
counted galaxies in annuli of 100 kpc, and also corrected for missing areas
due to large, extended foreground galaxies or stars.
As seen from this figure, the distribution of galaxies seems to fall off
outwards, but it is not completely clear whether the quasars lie at the 
centre of the distributions. 

\section{Comparison with other studies}

\label{section:s6}

\begin{figure*}
\begin{minipage}{14.1cm}
\psfig{file=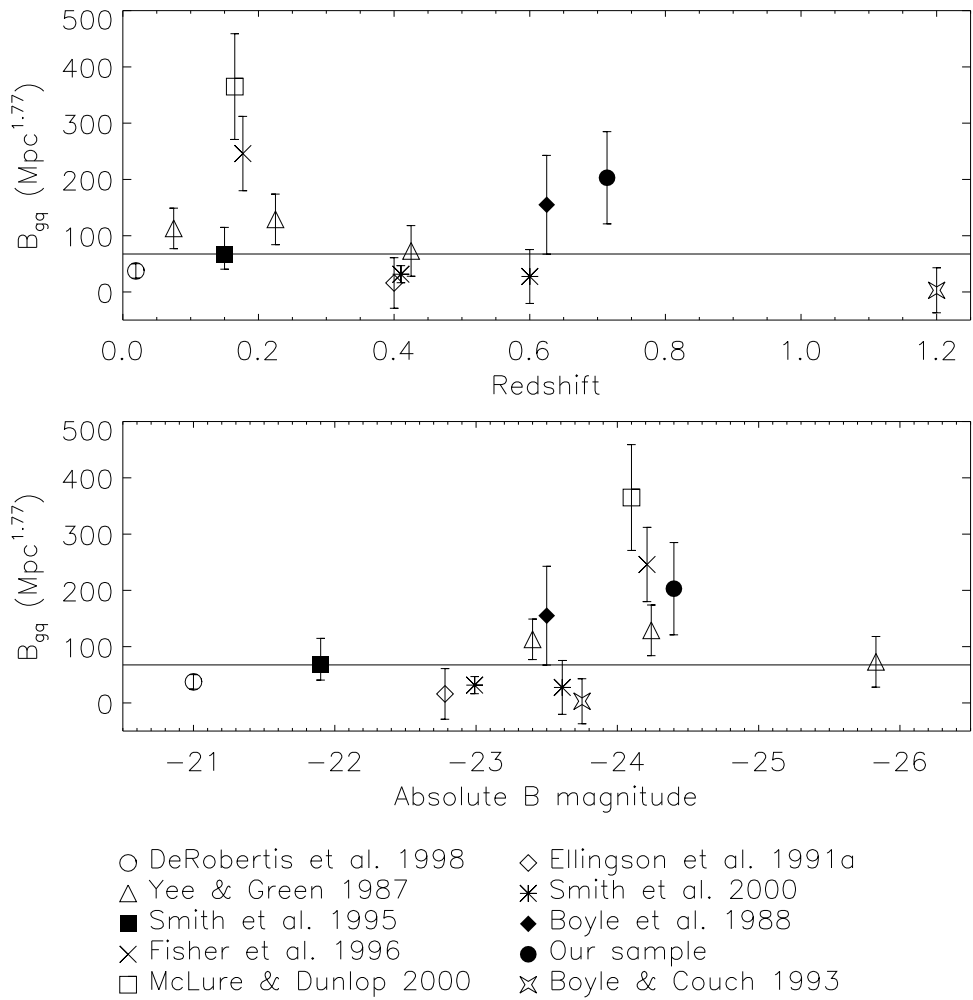}
\caption{RQQ environment as a function of redshift and $M_{B}$. The horizontal line
indicates the amplitude of the local galaxy--galaxy correlation
function, $B_{\rm gg}=67.5$ Mpc$^{1.77}$ \protect\cite{dp83}.}
\label{fig:fig7}
\end{minipage}
\end{figure*}

In this section we compare our result with those obtained
for other samples of RQQs.
A compilation of results from the literature is shown in 
Fig.~\ref{fig:fig7}, where we have plotted mean clustering
amplitudes as a function of median redshift and absolute 
quasar $B$
magnitude.

The lowest redshift point in Fig.~\ref{fig:fig7} is the mean 
amplitude found for 33 CfA Seyfert galaxies 
(both type 1 and 2) by DeRobertis et al.\ \shortcite{dyh98}, 
$B_{\rm gq}=37.4\pm12.9$ Mpc$^{1.77}$. 
DeRobertis et al.\ compute mean
clustering amplitudes separately for Seyfert~1 and Seyfert~2
galaxies, but these are consistent with each other in the
case where the counting radius is 0.5 Mpc. Furthermore, 
Seyfert~2 galaxies are believed to contain hidden Seyfert~1 nuclei 
(e.g.\ Antonucci 1993), so we have chosen to use the
mean for all Seyferts.
Seyfert~1 galaxies are the low-luminosity counterparts 
of RQQs, so we 
have assumed a mean nuclear luminosity of $M_{B}=-21$ for this sample.
The median redshift is 0.0192 (from table~1, DeRobertis, Hayhoe \& Yee (1998)).

Yee \& Green \shortcite{yg87} found clustering amplitudes
for BQS quasars in three redshift ranges, 0.00--0.15 (17
quasars), 0.15--0.30 (16) and 0.30--0.50 (7).
They list the ratio of the mean amplitudes to the amplitude 
of the galaxy--galaxy correlation function, 
$B_{\rm gq}/B_{\rm gg}$. Assuming that $B_{\rm gg}=67.5$ Mpc$^{1.77}$ \cite{dp83}, 
we find that $B_{\rm gq}$ is 113$\pm$36, 129$\pm$45 and 73$\pm$45 Mpc$^{1.77}$
in the three redshift ranges, respectively. 
To find the median absolute magnitude in the three 
sub-samples, we converted the apparent $r$ magnitudes listed by 
Green \& Yee \shortcite{gy84} and Yee et al.\ \shortcite{ygs86} 
to $R$, and thereafter to $B$ as described in 
Section~\ref{section:s3}. We found
median absolute $B$ magnitudes of $-$23.4, $-$24.24 and
$-$25.83.

Smith et al.\ \shortcite{sbm95} report a mean amplitude of
$B_{\rm gq}/B_{\rm gg} = 1.0^{+0.7}_{-0.4}$ (2$\sigma$ errors)
for their sample of 169 $z \leq 0.3$ X-ray selected quasars
with median redshift 0.15.
The quasar $M_{V}$ magnitudes in this sample span a wide range, from 
$\approx -27$ to $\approx -20$, 
but since there is no significant difference
between the amplitudes of the angular correlation 
function for their high- and 
low luminosity sub-samples, we have used the peak 
of the magnitude distribution,  
$M_{V}=-22.2$, converted to $M_{B}=-21.9$ assuming $B-V=0.3$.

Fisher et al.\ \shortcite{fisher96} find an average 
$B_{\rm gq}$ of 72$\pm$20 [h$^{-1}$Mpc]$^{1.77}$ for a sample 
of 14 optically bright RQQs, with a
median redshift of 0.24, and a median
$M_{V}$ of $-$23.5 ($h=1$). Assuming $B-V=0.3$ and $h=0.5$, 
we transformed to $B_{\rm gq}=246 \pm 66$ Mpc$^{1.77}$ 
and $M_{B}=-24.71$.
The RQQs in Fisher et al's sample are included in McLure \& Dunlop's 
\shortcite{md00}
sample, along with seven other sources. McLure \& Dunlop 
find a mean $B_{\rm gq}$ of 326$\pm$94 Mpc$^{1.77}$
and their sample has a median redshift of 0.165, 
and a median $M_V=-24.4$, which was converted
to $M_B=-24.1$.

At median $z=0.4$, we find the 32 RQQs in Ellingson 
et al's \shortcite{eyg91a} sample. The  mean amplitude is
16$\pm$45 Mpc$^{1.77}$, and we 
transformed the apparent quasar $r$ magnitudes (given in table~2, 
Ellingson et al.\ \shortcite{eyg91b})
to absolute $B$ magnitudes as described in Section~\ref{section:s3},
and found a median $M_{B}=-22.78$.
 
The 82 X-ray selected quasars studied by Smith et al.\ \shortcite{sbm00} cover 
the redshift range 0.3--0.7, and 
in their figure~8, $B_{\rm gq}$ is plotted as
a function of $M_{V}-5\log h$. By reading pairs of 
$B_{\rm gq}$ and $M_{V}-5\log h$ off the axis of this figure, we estimate that 
the $z<0.5$ and the $z>0.5$ sub-samples have median 
$M_{V}-5\log h \approx -21.8$ and 
$\approx -22.4$, respectively. These were converted to $M_{B}=-23.00$ and $-$23.61
assuming $h=0.5$ and $B-V=0.3$. We found the mean 
amplitudes in these two redshift intervals to be approximately
6.13$\pm$2.95 and 4.52$\pm$7.88 [h$^{-1}$Mpc]$^{1.77}$. (We estimate that
we can read off $M_{V}-5\log h$ and $B_{\rm gq}$
with an accuracy of $\pm 0.1$ and $\pm 5$ 
[h$^{-1}$Mpc$^{1.77}$], respectively.)
Smith et al.\ use a definition of $B_{\rm gq}$ which assumes
that the clustering is stable in co-moving coordinates,
whereas the definition by Longair \& Seldner \shortcite{ls79}
that we employ, assumes stable clustering in proper coordinates.
We therefore transformed Smith et al's 
amplitudes to our convention by multiplying with 
$\left(1+z\right)^{3-\gamma}$, assuming that $\gamma=1.77$, and that
$z=0.4$ and $z=0.6$ for the two sub-samples.
Thereafter, we transformed to the $h=0.5$ case, and got mean 
amplitudes of 31.6$\pm$15.2 and 27.5$\pm$47.9 Mpc$^{1.77}$. 

The next point along the redshift axis represents the results reported
by Boyle, Shanks \& Yee \shortcite{bsy88} for eight BFSP quasars in the 
redshift range $0.55<z<0.7$, with mean 
$B_{\rm gq}/B_{\rm gg} = 2.3\pm1.3$. We have 
assumed $M_{B}=-23.5$ and a mean redshift
of 0.625 for these quasars, since no other information is given. 

Our own sample is plotted with a mean $B_{\rm gq}=210\pm82$
(using Model~3), and a median redshift and $M_B$ of 
0.714 and $-$24.40.

Finally, the high-$z$ point corresponds to 
the work by Boyle \& Couch \shortcite{bc93}, who find a mean 
$B_{\rm gq}=3\pm40$ Mpc$^{1.77}$. The median $M_B$ is $-$23.75,
and we have assumed a mean redshift of 1.2.

There are several aspects about the quasar phenomenon, such as formation,
fuelling and evolution, that we can learn about by looking
at their environments as a function of redshift and AGN luminosity.
But this has been difficult to do, both because there appears to be a wide range
in richnesses within samples, and also because it requires good coverage of 
the redshift-luminosity plane. A figure like Fig.~\ref{fig:fig7}
might help, since looking at the mean $B_{\rm gq}$ values over a wide
range in redshift and luminosity is more meaningful than relying
on amplitudes found in single fields. The drawback is of cause that 
Fig.~\ref{fig:fig7} includes results from different studies with 
differing selection criteria, observations, analysis etc., and therefore
contains uncontrollable selection effects and biasses. We are
therefore careful with interpretations, and merely look at it as 
a useful way of comparing different surveys to see how they 
agree or disagree with each other as we move along the redshift and the 
luminosity axis. 

Approximately half of the data points are significantly higher than the 
galaxy--galaxy clustering amplitude 
of $B_{\rm gg}=67.5$ Mpc$^{1.77}$, whereas the other half is consistent
with, or less than, $B_{\rm gg}$.
The samples from the large cross-correlation studies 
(Smith et al.\ 1995; 2000; Boyle \& Couch 1993) belong to the latter, 
along with Ellingson et al's sample, Yee \& Green's high-$z$ sample
and the Seyfert sample of DeRobertis et al. 
The Seyfert sample is perhaps not so difficult to explain if we consider
that spiral galaxies prefer field-like environments. If we further
assume that low-luminosity quasars are mostly hosted by spiral galaxies,
this might also explain Ellingson et al's result, but it does not
explain why the mean amplitude of Yee \& Green's bright, high-$z$ sample
is so low.

It is however noteworthy that the datapoints which agree with a galaxy
environment richer than that of field galaxies are clustered around
$M_B\approx -24$. The exceptions are Boyle \& Couch's sample and Smith et al's (2000)
$0.5 < z < 0.7$ sample, which have 
$B_{\rm gq} < B_{\rm gg}$. This may lead us to think that in these
two studies, perhaps the galaxy excess (if present) was not detected.
Their observations were done with filters that targets emission on the
short-wavelength side of the 4000 {\AA} break at the quasar redshift,
and we found in Section~\ref{section:s5} that the clustering signal
in fields with a clear excess is weaker shortward of 4000 {\AA}. 
Boyle \& Couch used $R$ and a wide $VR$ 
passband (a combination of Johnson $V$ and Kron-Cousins $R$), and at $z\approx 1.2$ 
both these filters target emission shortward of the 4000 {\AA} break, 
and the sensitivity to early-type galaxies
might not be particularly good. Also, the limiting magnitude,
$R=23$, might be somewhat too shallow in order to reach sufficiently 
faint into the luminosity function at these redshifts.
Smith et al.\ \shortcite{sbm00} used $V$-band, also corresponding
to emission shortward of the 4000 {\AA} break, at least at $z>0.4$ where the bulk 
of their quasars are.

Most of the clustering studies similar to our (in terms of method and 
analysis) seem to agree fairly well that RQQ environments are 
different from field-like environments, both at high and at low redshifts.
Yee \& Green, Fisher et al.\ and McLure \& Dunlop, looking at 
bright quasars
at low redshifts, get clustering amplitudes consistent with ours, 
although Yee \& Green's is a bit lower than the other two.
Boyle et al's mean amplitude for BFSP quasars at $0.55 < z < 0.7$
also agree with ours at $z\approx 0.7$ and Fisher et al's at $z\approx 0.2$,
so this raises the possibility that in fact there is little change in 
RQQ environments from $z\approx 0.2$ to $z\approx 0.7$--0.8. This is 
in agreement with recent studies that also find no changes in RLQ environments with 
redshift (Paper~I; McLure \& Dunlop 2000).

In light of the above, and considering the similarities in method and
analysis, the low amplitude from Ellingson et al's work is difficult
to explain. Since our quasar sample cover a wide range in $B$
luminosity, there are several quasars at similar luminosities 
to Ellingson et al's quasars. The six quasars at $M_B>-23.5$ in our
sample have a mean $B_{\rm gq}=237 \pm 78$ (Model~3), which is
very different from Ellingson et al's 16$\pm$45 Mpc$^{1.77}$. 
The remaining 14 quasars at $M_B < -23.5$ have a mean amplitude of 196$\pm$110
Mpc$^{1.77}$ (Model~3), so we find no difference between high- and low-luminosity
sub-samples. Since the six quasars at $M_B > -23.5$ have a median redshift of 
0.720, i.e.\ higher than Ellingson et al's $z=0.4$, it might seem as though the 
environmental richness change with redshift for the lower luminosity quasars,
but not for the high luminosity ones. 
This is perhaps somewhat odd since we have no reason to 
believe that our amplitudes have a systematic error relative to Ellingson et al's. 

In Paper~I, we found that the $B_{\rm gq}$'s for our
RLQ fields lay in the same range as those found for RLQs by Ellingson 
et al., and the mean amplitudes were consistent with each other.
Furthermore, for the RLQ MRC 0405$-$123, Yee \& Ellingson \shortcite{ye93}
found $B_{\rm gq}=905\pm277$, whereas we found 579$\pm$179 Mpc$^{1.77}$, i.e.\
slightly lower, but still in agreement with each other. So it   
seems unlikely that our amplitudes are systematically different from 
Ellingson et al's. One possible explanation for the discrepancy may be
the inhomogeneity of Ellingson et al's sample. Their sample was chosen from the 
Hewitt-Burbidge catalogue \cite{hb87}, which is an incomplete compilation of 
quasars, but nevertheless the best catalogue available at that time. 
There could be some subtle bias in this sample that causes the highest $B_{\rm gq}$
fields to be excluded. E.g.\ if we clip out the five highest $B_{\rm gq}$'s in our 
sample, we get a similar result as Ellingson et al., most RQQs having $B_{\rm gq}$ 
less than 300 Mpc$^{1.77}$ and a mean consistent with their. 

We have seen that, whereas some surveys find that RQQs
live in field-like environments, others agree with our survey 
finding a richer than average galaxy density around RQQs. 
But the question whether the disagreement is due to selection 
effects caused by redshift and/or luminosity, appears difficult
to answer.


\section{A direct comparison with RLQ environments}

\label{section:s7}

In Paper~I, we investigated the galaxy environments around 
21 steep-spectrum RLQs. 
The current sample of RQQs matches the RLQ sample in redshift and
$B$ magnitude, and both samples were observed with the same 
telescope and analysed in the same manner.
This offer us a direct and internally consistent way
of comparing the galaxy environments around RLQs and RQQs
at the same epoch and with similar AGN luminosities. 

\begin{figure}
\psfig{file=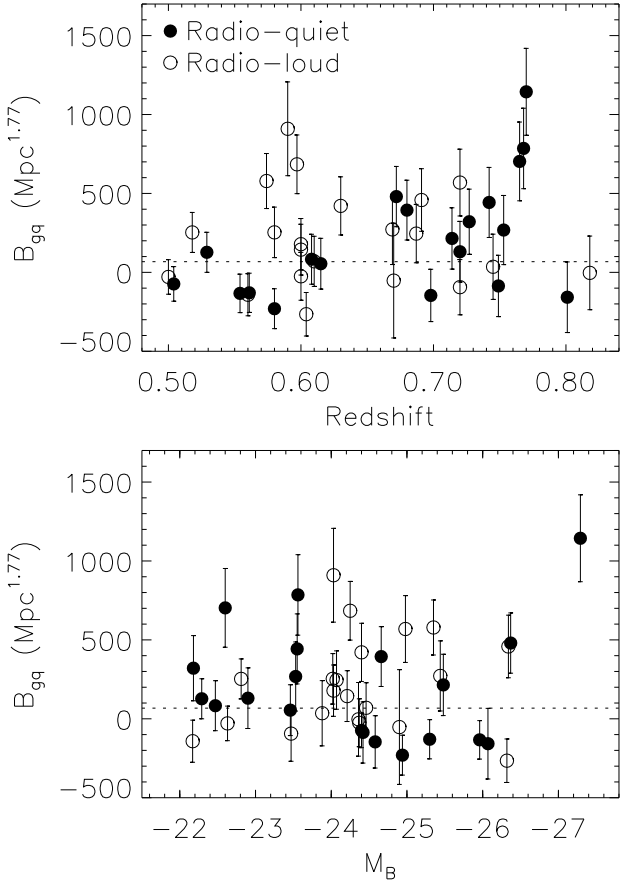}
\caption{$B_{\rm gq}$ as a function of redshift and quasar absolute 
magnitude for RLQs and RQQs.
The dotted lines refer to the value of the amplitude of 
the local galaxy--galaxy 
correlation function, $B_{\rm gg} = 67.5$ Mpc$^{1.77}$ \protect\cite{dp83}.}
\label{fig:fig8}
\end{figure}

In Fig.~\ref{fig:fig8}, we show $B_{\rm gq}$ as function of 
$z$ and $M_B$ for both samples (using Model~3 for the background counts),
where it can be seen that they span the same wide range in 
environmental richness.
The mean clustering amplitude for the RLQs is
213$\pm$66 (Model~3), and redoing the RLQ fields 
using Model~1 and 2 for the background counts, 
the mean $B_{\rm gq}$'s become 214$\pm$83 and 224$\pm$55 
Mpc$^{1.77}$, respectively.
We thus find that the mean clustering amplitudes for the RLQ and the 
RQQ samples are practically indistinguishable, i.e.\ {\em on average, 
there is no difference in the galaxy environments on 0.5 Mpc scales for the 
RQQs and the RLQs}. 

The mean and the median $B_{\rm gq}$'s for the two quasar samples are 
summarized in Table~\ref{table:table6}. The median $B_{\rm gq}$'s are seen to 
be generally lower than the mean values, but this is not very significant. 
The intrinsic scatter seems to be smaller for the RLQs, and is more
comparable to $\Delta B_{\rm gq}$ than for the RQQs. 
A KS test gives a significance level of 0.72 for the null hypothesis that
the two samples are drawn from the same distribution, i.e.\ the two samples
are consistent with being drawn from the same distribution.

This result is contrary to the traditional picture that the 
environments of RLQs and 
RQQs differ, i.e.\ that RQQs are found in systematically 
poorer environments than RLQs. It was reported in the similar study
by Ellingson et al.\ \shortcite{eyg91a} that RLQs inhabited poorer environments
than RQQs at $z\sim 0.4$--0.5. 
However, at $z \approx 0.2$, Fisher et al.\ \shortcite{fisher96}
and McLure \& Dunlop \shortcite{md00} compare 
samples of RLQs and RQQs matched in
redshift and luminosity, and find no significant differences, hence our result at $0.5 \leq z \leq 0.8$ 
is consistent with these findings at low redshifts. 
Also at higher redshifts, $z\approx1.1$, Hutchings et al.\ \shortcite{hcj95} find
evidence for similar environments around RLQs and RQQs, 
compact groups or clusters of star-forming galaxies.
It thus appears that both RLQs and RQQs 
at low to intermediate redshifts on average 
occupy the same type of environment, consistent with poorer galaxy clusters,
and that the environments 
of, at least powerful, RLQs and RQQs do not change with redshift, 
possibly not even out to $z \sim 1$.

Our result is also consistent with the findings that at low redshifts 
powerful RLQs and RQQs appear to be hosted by luminous 
galaxies with luminosities above the break in the luminosity function 
(e.g.\ Dunlop et al.\ 1993; McLeod \& Rieke 1994; 
Taylor et al.\ 1996; Bahcall et al.\ 1997; Boyce et al.\ 1998).
Luminous early-type galaxies are known to prefer rich environments, and it thus
follows that powerful quasars hosted by these galaxies also appear in rich environments.
There are some evidence that the hosts of powerful RQQs are 
massive ellipticals, but the results are still somewhat inconclusive.
Bahcall et al.\ \shortcite{bahcall97} find more RQQs in elliptical hosts
than in spiral hosts, but note that the quasars also appear in interacting systems and
spiral galaxies.
McLure et al.\ \shortcite{mclure99} find that all $M_R < -24$ RQQs 
in their sample are hosted by massive ellipticals and  
Hughes et al.\ \shortcite{hughes00} present off-nuclear spectra of luminous AGN, and identify 
the 4000 {\AA} break in the stellar component of the host in many cases. 

\begin{table*}
\begin{minipage}{14cm}

\caption{Mean and median clustering amplitudes for 21 RLQs 
and 20 RQQs. 
Model~1--3 correspond to different ways of correcting for background galaxies. The intrinsic
scatter in the $B_{\rm gq}$ distribution is denoted $\sigma_{\rm int}$ and was calculated as 
$(\sigma_{B_{\rm gq}}^{2} + \Delta B_{\rm gq}^{2})^{1/2}$,
where $\sigma_{B_{\rm gq}}$ is the standard deviation of 
the $B_{\rm gq}$ distribution and $\Delta B_{\rm gq}$ is the
measurement error.}

\begin{tabular}{llllllllll}
 & \multicolumn{3}{c}{Model 1} & \multicolumn{2}{c}{Model 2} & \multicolumn{3}{c}{Model 3} \\
Sample & Mean & Median & $\sigma_{\rm int}$ & Mean & Median & $\sigma_{\rm int}$ & Mean & Median & $\sigma_{\rm int}$ \\
 & & & & & & & & & \\
RQQ & 336$\pm$77 & 229 & 387$\pm$54 & 212$\pm$74 & 151 & 276$\pm$72 & 210$\pm$82 & 131 & 315$\pm$73 \\ 
RLQ & 214$\pm$83 & 264 & 332$\pm$75 & 224$\pm$55 & 165 & 172$\pm$86 & 213$\pm$66 & 178 & 236$\pm$76 \\
\end{tabular}
\end{minipage}
\label{table:table6}
\end{table*}

For completeness, we mention that 
if we exclude the three quasars with radio powers $> 1.3 \times 10^{25}$ W\,Hz$^{-1}$ 
(the two BQS quasars and LBQS 2348$+$0210), the mean becomes 191$\pm$73 
Mpc$^{1.77}$ (Model~3), 
so there is little change, and the mean is still consistent with the mean 
$B_{\rm gq}$ for the RLQs. This is also the case when we exclude the two BQS quasars on 
the basis that the BQS suffers from incompleteness, the mean being 177$\pm$71 Mpc$^{1.77}$.
The mean amplitude for the sample when the seven radio-detected
quasars are excluded is 178$\pm$82 Mpc$^{1.77}$.


\section{Discussion}

\label{section:s8}

In Paper~I, we found that the clustering amplitudes in RLQ
fields seemed
to correlate with radio luminosity instead of redshift.
Extrapolating this to the RQQs with radio luminosities two--three
orders of magnitude lower, one 
might expect them to be sited in poorer environments than the 
RLQs.
Instead we find that, on average, the environments around RLQs and
RQQs are indistinguishable. This suggests that the process that 
determines the radio loudness is not dependent on the environment
on Mpc scales, but may instead be found in the central regions
of the host galaxy. 
This is consistent with the correlation found by Rawlings
\& Saunders \shortcite{rs91} between narrow-line luminosity
and radio luminosity in radio galaxies, and the radio-optical
correlation in steep-spectrum RLQs found by Serjeant
et al.\ \shortcite{serjeant98}. These correlations link the
production of optical emission lines in the nuclear regions
to the production of large-scale radio emission, implying
that the fuelling of radio jets is linked to accretion 
onto the black hole.

If, e.g.\ as a result of galaxy interactions and mergers,
a RLQ is born, the environment into which the source expands
may affect its radio luminosity, e.g.\ enhancing it
\cite{ba96}, and we observe a weak correlation between
$B_{\rm gq}$ and radio luminosity (Paper~I), but the presence 
of the radio loudness itself 
would be independent of environment. 
If a RQQ is born, we would perhaps expect a correlation between 
quasar optical luminosity and $B_{\rm gq}$ (also for RLQs), but
this would be harder to detect, both because the nuclear 
optical luminosities probe much smaller regions than the radio 
luminosity does, and because the optical luminosity may vary.

We expect a relation between environmental richness and quasar 
optical luminosity
since the luminosity created by accretion onto the black hole scales with the 
black hole mass. According to the Kormendy \& Richstone \shortcite{kr95}
and the Magorrian et al.\ \shortcite{magorrian98} relations, the mass of a black hole
scales with the luminosity (or the mass) of the spheroidal component
of the host galaxy. (See also Ferrarese \& Merritt (2000) who find a tight
relation between black hole mass and velocity dispersions of bulges.) 
Early-type galaxies, being 
bulge-dominated, and likely to reside in rich environments, are therefore expected to host
the most massive black holes, and thus the most luminous quasars. 
McLure \& Dunlop \shortcite{md00} looked at $B_{\rm gq}$ vs nuclear luminosity 
as found from the Magorrian et al.\ relation, but found that the correlation was 
not statistically significant. 

Being sited in equally rich environments, RLQs and RQQs are
equally replenished by fuel from their surroundings, so both
quasar populations would survive for similar amounts of time. 
This implies that the fraction of radio-loud quasars would be 
independent of redshift, and agrees well with Goldschmidt et al.\
\shortcite{goldschmidt99} who find that the radio-loud 
fraction depends on optical luminosity, not redshift.

There does not appear to be evidence for an epoch dependence in $B_{\rm gq}$
between $z\approx 0.2$ and $\approx 0.8$. It is however not unthinkable that 
there could be one for RQQs, but at higher redshifts, $z\sim 1$--2. 
This is because hosts of powerful RQQs change with redshift;
at low-$z$, they are giant elliptical galaxies with $L>L^{*}$ (the less powerful ones have 
spiral hosts), whereas at high redshifts they are $\sim L^{*}$ galaxies (e.g.\ Hutchings 1995b; 
Ridgway et al.\ 2000; Kukula et al.\ 2000). 
Radio galaxies do not change over the whole interval $z\sim2$ to 0 
(Lacy, Bunker \& Ridgway 2000), 
so according to radio-loud unified schemes, RLQ hosts should behave in the same manner.
Arag{\'o}n-Salamanca, Ellis \& O'Brien \shortcite{aeb96} do in fact find a significant excess
around RLQs at $z\approx 2.5$ that is not seen around RQQs at the same redshift
(four RLQs and six RQQs).

The change in the hosts of RQQs with redshift is probably related to the 
huge increase in their space density between $z\sim 0-1$ and $z\sim 2-3$. In a model for
the formation and evolution of quasars, Kauffmann \& Haehnelt \shortcite{kh00} link their 
evolution to the hierarchical build-up of galaxies, and predict that quasars at a 
given luminosity should be found in progressively less luminous
host galaxies towards higher redshifts. 
But this explanation cannot work for the radio-louds.
The fact that the hosts of the radio-louds 
do not appear to evolve strongly may be because there is a 
strong link between the mass of the black hole and the ability 
to produce powerful jets (Lacy, Ridgway \& Trentham 2000; Laor 
2000). Thus only objects whose black holes (and therefore
hosts) are already massive at high redshifts can be radio
loud.

In principle, we can also quote a lower limit to the 
co-moving space density of Abell 1--2 clusters at $z\sim0.7$.
The space density of quasars at $z=0.7$ is
$\sim 3 \times 10^{-6}$ Mpc$^{-3}$\footnote{Found by 
integrating Boyle et al's \shortcite{boyle00} quasar luminosity function between the 
absolute magnitude limits of our survey. We used the expression
$\frac{\Phi(M_{B},z)}{\Phi(M^{*}_{B})}=[10^{-0.4[(\alpha-1)(M_{B}-M^{*}_{B}(z))]}+10^{-0.4[(\beta-1)(M_{B}-M^{*}_{B}(z))]}]^{-1}$, which is different from that given in the paper, since we 
suspect that a misprint may have occurred.} \cite{boyle00}, and since 
$\approx 10$ per cent of the quasars in our sample live in
Abell 1--2 clusters, this would correspond to 
$>$3$\times$10$^{-7}$ Abell 1--2 clusters per Mpc$^{3}$
at $z\sim 0.7$. This is potentially interesting for
cosmology and may even constrain $\Omega_{0}$, because 
predictions of cluster number densities at high redshifts
are highly model dependent \cite{lilje92}.

It is also interesting to note that this can be used to 
set a lower limit to the duty cycle of quasars. Normally,
only upper limits are given based on e.g.\ central
black hole masses in Seyferts.
The duty cycle of quasars can be thought of as the 
probability for a black hole to accrete, or the ratio of 
active to dormant + active quasars at a certain epoch.
We know from our study that the fraction of active quasars living
in Abell 1--2 clusters at $z\approx0.7$ is $\approx 0.10$, and we
also know that the number of clusters hosting a black hole (both dormant 
and active) must be less than the number of Abell 1--2 
clusters at the present epoch. 
If we assume that the typical X-ray luminosity of such clusters is 
10$^{43}$ erg\,s$^{-1}$, the local cluster density is $\approx 2\times$10$^{-5}$
\cite{rosati98}, and thus the lower limit to the quasar duty cycle is
$\approx 0.1$$\times$3$\times$10$^{-6}$/2$\times$10$^{-5}$, i.e. about 1 per cent. 
This only applies to RQQs living in Abell 1--2 clusters,
and the present-day duty cycle of such systems is probably
very low. The continous accretion phase cannot last very
long, and at some point it stops, perhaps when the cluster 
virializes.


\section{Conclusions}

\label{section:s9}

The RQQs studied in this paper live in a diversity of 
environments, from field-like environments to galaxy clusters  
as rich as Abell class 1--2. 
On average, the sample has a clustering amplitude 
corresponding to 
groups or poorer clusters of Abell class $\approx 0$, 
and is {\em indistinguisable} from the mean amplitude found 
for a sample of
RLQs matched in redshift and AGN luminosity. 

Our finding that the average environment around RLQs and RQQs 
at $0.5 \leq z \leq 0.8$ is statistically indistinguishable
agrees with what is found for powerful quasars at 
$z\approx0.2$ by Fisher et al.\ \shortcite{fisher96} and 
McLure \& Dunlop \shortcite{md00}. 
There is agreement also in terms of the 
amount of clustering, suggesting that quasar environments,
for both RLQs and RQQs, do 
not undergo much evolution between $z\approx0.2$ and 
$\approx0.8$. These results are consistent 
with host galaxy studies finding that both powerful RLQs and 
RQQs
live in massive, elliptical galaxies 
(e.g.\ Dunlop et al.\ 1993;
Bahcall et al.\ 1997; McLure et al.\ 1999).
We therefore believe that the consensus that 
RQQs occupy systematically poorer environments than RLQs 
should be put under doubt.

Since our RQQs on average appear in cluster environments 
they are most likely biased with 
respect to galaxies. Galaxy clusters are also biased 
(e.g.\ Lilje \& Efstathiou 1988; Croft, Dalton \& Efstathiou 1999), 
so the RQQs in clusters must have a bias factor that is greater 
than that for normal galaxies. This suggests that RQQs can not be
used as unbiased tracers of galaxies when studying large scale 
stucture using quasars surveys.  

That RLQs and RQQs on average live in equally rich environments
implies that the origin of the radio loudness is not
linked to the environments on Mpc scales, but may instead be
buried within the central parts of the host galaxy.
This is consistent with the suggestion that the fuelling of 
radio jets is linked to accretion onto a central black 
hole (Rawlins \& Saunders 1991; Serjeant et al.\ 1998).

A colour-magnitude diagram of the 
richest RQQ fields in this study shows hints of a red 
sequence at the expected colour of $z = 0.7$--0.8 
early-type galaxies. With spectroscopy, these fields will 
likely turn out to contain galaxy clusters at $z=0.7$--0.8.
RLQs and radio galaxies are often used to find and study 
galaxy clusters at higher redshifts, and our result that 
RLQs and RQQs inhabit similar environment at $z \sim 0.7$--0.8 
should therefore increase the pool of likely cluster candidates 
at high redshifts. This may be important for theories of 
galaxy formation, since it allows us to find and study 
high redshift galaxy clusters that are not selected on the 
basis of extreme optical or X-ray luminosity. 


\section*{Acknowledgements}

We are grateful to the staff at the NOT for help during the observations, 
to D. Burstein for providing maps of galactic extinction in electronic 
form, and to N. Kaiser for the {\sc imcat} software. 
MW wishes to thank H. Dahle for valuable help with {\sc imcat}, L. Borgonovo for 
discussions,  
and also D. Koo, C. Willmer and P. Guhathakurta for discussions during a
visit to UCSC. PBL and MW acknowledges travel support from the Norwegian
Research Council, and MW also acknowledges the STINT
Exchange Program Stockholm Observatory-Astronomy/UC Santa Cruz.
ML acknowledges support from NSF grants AST-98-02791 and AST-98-02732,
and conducted his work under the auspices of the U.S. Department of Energy at the
University of California Lawrence Livermore National Laboratory under 
contract No.\ W-7405-Eng-48.

This research is based on observations made with the Nordic Optical 
Telescope which is 
is operated on the island of La~Palma jointly by Denmark, Finland,
Iceland, Norway and Sweden, in the Spanish Observatorio del Roque de 
los Muchachos of the Instituto de Astrofisica de Canarias. 
The data presented here have in part been taken using ALFOSC, which is owned by
the Instituto de Astrofisica de Andalucia (IAA) and operated at the 
Nordic Optical Telescope under agreement between IAA and NBIfAFG of the
University of Copenhagen.

{\sc iraf} is distributed by the National Optical Astronomy 
Observatories, which are operated by the Association of Universities 
for Research in Astronomy, Inc., under cooperative agreement with the 
National Science Foundation. This research has made use of the 
NASA/IPAC extragalactic database (NED) which is operated by the Jet 
Propulsion Laboratory, California Institute of Technology,
under contract with the National Aeronautics and Space Administration. 


\end{document}